\documentclass[11pt]{article}
\usepackage{authblk}
\usepackage[a4paper]{geometry}
\usepackage{graphicx}
\usepackage[textwidth=8em,textsize=small]{todonotes}
\usepackage{amsmath}
\usepackage{hyperref}
\usepackage{textcomp}
\usepackage{epstopdf}
\usepackage{setspace}
\usepackage{multirow}
\usepackage{tabularx}
\newcolumntype{Y}{>{\centering\arraybackslash}X}  
\usepackage{longtable} 
\usepackage{pdflscape}
\usepackage{bm} 
\usepackage{gensymb}
\usepackage{csquotes}

\makeatletter 
\renewcommand\@biblabel[1]{#1.} 
\makeatother

\title{First principles study of the surface of silica and sodium silicate glasses}

\author{Zhen Zhang$^{(1),(2)}$}
\author{Simona Ispas$^{(1)}$\thanks{Corresponding author: \texttt{simona.ispas@umontpellier.fr}}}
\author{Walter Kob$^{(1)}$}
\affil{$^{(1)}$ Laboratoire Charles Coulomb,
	University of Montpellier, CNRS, \\
	F-34095 Montpellier, France\\
	$^{(2)}$ Center for Alloy Innovation and Design, State Key Laboratory for Mechanical Behavior of Materials, Xi'an Jiaotong University, Xi'an 710049, ChinaI
}

\begin{document}

\maketitle

\begin{abstract}  
We use \textit{ab initio} molecular dynamics simulations to investigate
the properties of the dry surface of pure silica and sodium silicate glasses. The
surface layers are defined based on the atomic distributions along the
direction ($z-$direction) perpendicular to the surfaces. We show that these
surfaces have a higher concentration of  dangling bonds as well as two-membered (2M)
rings than the bulk samples. Increasing  concentration of  Na$_2$O reduces
the proportion of structural defects. 
 From the vibrational density of states, one concludes  that 2M rings have a unique vibrational signature
at a frequency $\approx850$~cm$^{-1}$, compatible with experimental
findings.
 We also find that, due to the presence of surfaces, the
atomic vibration in the $z-$direction is softer than for the two other
directions. The electronic density of states shows clear the
differences between the surface and interior and we can attribute these  to specific
structural units. Finally, the analysis of the electron localization
function allows to get insight  on the influence of local structure and the
presence of Na on the nature of chemical bonding in the glasses.

\end{abstract}


\section{Introduction}

Developing glasses with tailored surface properties is
an important step in many types of applications such as
the production of ultra-thin flexible displays and energy
efficient windows, catalysis technology, electronics, and biomaterials
\cite{Pantano1989,bocko1991surface,bach1997advanced,zhuravlev_surface_2000,pugliara2016assessing,dey2016cleaning,zheng2019protein}.
Among the problems one faces in the design of glasses with specific
properties are the presence of surface defects which can lead to a
dramatical drop of the mechanical strength or a strong alternation of the
chemical reactivity of the samples. In spite of the considerable number
of experimental and computational studies that have probed the surface
properties of silicate glasses, we still lack  atomistic insight
how the glass composition affects the structure of such surfaces or the
concentration of defects, or how the local structure of the surface
influences the spectroscopic and electronic properties.
Note that when we discuss here glass surfaces, we refer to surfaces that
have been obtained by cooling the glass-former from the melt, i.e., we
do not consider the case in which the surface is produce by a fracture
process~\cite{anderson_fracture_2017,zhang_surfcl_2020}.

Experimental techniques such as low-energy ion scattering (LEIS)
spectroscopy, X-ray photoelectron spectroscopy (XPS), or atomic force
microscopy (AFM) have provided information on surface composition and its
microstructure~\cite{kelso_comparison_1983,almeida_low-energy_2014,almeida_low-energy_2016,cushman2018low,radlein1997atomic,poggemann2001direct,poggemann2003direct,frischat2004nanostructure}.
The LEIS technique has, e.g., allowed to demonstrate that
melt-formed surfaces of binary oxide glasses are depleted of the
modifier atoms which seem to evaporate when the sample is still
in the liquid state~\cite{almeida_low-energy_2016}. AFM measurements have allowed
to probe the structural features of glass surfaces with atomic
resolution, and thus to obtain structural information such as interatomic
distances and grouping of atoms, but the nature of the defects could not be
determined~\cite{radlein1997atomic,poggemann2001direct,poggemann2003direct,frischat2004nanostructure}. 
Information on the surface structure can also be
obtained from spectroscopic techniques such as infrared (IR)
spectroscopy, nuclear magnetic resonance (NMR), extended X-ray absorption fine structure (EXAFS) or electron paramagnetic resonance  (EPR)~\cite{radzig2007point,berruyer2017three}. However, in order to obtain from such techniques
information about the structural properties of the surface it is
usually necessary to make a hypothesis on the nature of the defects
and/or to combine spectral, kinetic and computational data, a task that is not
straightforward at all~\cite{comas2017understanding,berruyer2017three}. 

Pure silica is the simplest silicate glass and because
of its importance in industrial and engineering
applications such as support medium for modern heterogeneous
catalysts and biomolecules it has been widely studied in the past
\cite{zhuravlev_surface_2000,varshneya2013fundamentals,rimola_silica_2013,tielens2019Characterization}.
Experimental as well as theoretical studies have given evidence that the
local structure of the outermost layer of silica surfaces consists of
SiOSi bridges (called siloxane bridges) and SiO$_4$ tetrahedra bearing one
or two OH groups~\cite{zhuravlev_surface_2000,rimola_silica_2013}. Using
appropriate heating and thermal treatment (above 700$^{\circ}$~C),
the concentration of these silanol groups can be reduced,
allowing to generate partially or even fully dehydroxylated silica
surfaces~\cite{morrow1976infrared,bunker1989infrared,grabbe1995strained,coperet2003homogeneous,sot2019fully}.
With the reduction of the surface hydroxylation, defective structures
are generated, in particular  strained two-membered (2M) rings, i.e.~two 
tetrahedra that share an edge. The presence of this
type of defect, completely absent in the bulk sample, was inferred from
the appearance of certain features in the IR spectra, namely
two bands at 888 and 908~cm$ ^{-1} $, and a shoulder at  932~cm$ ^{-1}
$~\cite{morrow1976infrared,bunker1989infrared,grabbe1995strained,ferrari1995reactions}.
These 2M rings are under high local stress and hence are
considered to be important reactive sites capable to favor the
functionalization of the surface as indicated by various experimental
studies~\cite{coperet2003homogeneous,comas2017understanding}.
Other experiments indicate the existence of further local
defects, such as under-coordinated silicon and non-bridging
oxygen atoms, but their concentration and
the way they modify the network are not
known~\cite{vaccaro2008luminescence,sot2019fully}.

These experimental efforts have been complemented
by computer simulation studies, pioneered by the
classical molecular (MD) simulations of Garofalini and
co-workers~\cite{garofalini_molecular_1983,feuston_topological_1989,garofalini_molecular_1990}.
Using various types of interaction potentials, the surfaces of
silica glasses were investigated in detail in order to identify the different
structural features and in particular the concentration of the mentioned
defects~\cite{wilson2000hydrolysis,roder_structure_2001,rarivomanantsoa_classical_2001,wang2003molecular1,du_molecular_2005,gonccalves2016molecular,rimsza_surface_2017,halbert2018modelling}.
Although most of these studies did indeed report a finite
concentration of the various local defect sites, the values
did not match well the experimental data, likely because of the
used protocol to generate the samples or the inaccuracies of the
interaction potential. Similar investigations have also been
carried out for surfaces of more complex glasses and it was found
that their structure differed significantly from the one of the bulk
system~\cite{garofalini1985differences,ren_surface_2017,garofalini2018simulations}.

Note that most effective force fields used to carry out these simulations
have been developed to describe the bulk properties of glasses. Therefore
it is far from obvious whether or not such  classical MD simulations are able to
give a quantitative correct description of the local structure of the
surface since the arrangement of the atoms is very different from the
one encountered in the bulk. This problem can be avoided by using an
\textit{ab initio} approach in which the forces are directly calculated
from the electronic degrees of freedom~\cite{kob_first-principles_2016}. This approach is
thus not only more reliable but in addition it also allows to determine
the electronic signatures of the main structural features of samples
with surfaces.

The goal of the present work is thus to provide a detailed description
of silicate glass surfaces in terms of their structural, vibrational, and
electronic properties and to probe how these properties depend on the
composition of the glass. The use of {\it ab initio} calculations will
allow us to extract the spectroscopic and electronic signatures of the
defective sites and in particular to understand how the presence of Na
atoms affect the various properties.

The reminder of the paper is organized as follows: The next section
gives details on the composition of the studied glasses, the protocol
used to generate samples with surfaces, as well as on the adopted
computational framework. In Sec.~\ref{sec:structure} we will present a description
of the structural properties of the surface domain and compare these to the ones
of the interior (bulk-like) one. Subsequently we will
show and discuss the  vibrational (in Sec.~\ref{sec:vibrations}) and
electronic (in Sec.~\ref{sec:electronic}) properties of the studied
compositions. Finally, the last section will summarize the conclusions
and give perspectives of the present work.

\section{Simulation details} \label{sec:sim}

In this present study, we have considered three glass-forming systems: Pure silica
(SiO$_2$) and two binary sodo-silicates,  Na$_2$O-5SiO$_2$  and
Na$_2$O-3SiO$_2$, denoted hereafter as NS5 and NS3, respectively. To
start we prepared a bulk liquid sample containing around 400 atoms
randomly placed in a cubic simulation box and carried out classical
molecular dynamics simulations  at relatively high temperatures
(3600~K for SiO$_2$ and at 3000~K for the two sodosilicates), using
periodic boundary conditions. The initial box side was chosen so
that the density coincides with that of the glass of the corresponding
composition at room temperature~\cite{bansal_handbook_1986}. 
More details can be found in Ref.~\cite{zhang_fracture_2020}.
The final
configurations of these classical simulations were then used as starting
points for the equilibration runs carried out within the framework of {\it
ab initio} molecular dynamics (AIMD) simulations at the same temperatures
and using the constant volume–constant temperature ($NVT$) ensemble. The
lengths of these AIMD trajectories were 12.2~ps, 15.6~ps, and 11.8~ps
for silica, NS5 and NS3, respectively, a time span that was sufficiently
long to completely equilibrate the samples.  More details (composition,
number of atoms, densities, box sizes of bulk samples) are
given in Table~\ref{tab: ab simu-parameters}.  For each composition,
two independent samples were prepared and the results presented in the
following sections are their averaged properties.

\begin{table*}[ht]
	\small
	\center
	\begin{tabular}{lcccccc}
		\hline
		&{\hspace{2mm}}  \#atoms {\hspace{2mm}} & Na$_2$O-mole\% {\hspace{2mm}} &{\hspace{2mm}}  $L_{\rm bulk}$~(\AA) {\hspace{2mm}} & {\hspace{2mm}} $\rho_{\rm bulk}$~(g/cm$^3$) {\hspace{2mm}} & {\hspace{2mm}} $T_0$~(K)  {\hspace{1mm}} & {\hspace{2mm}} $T_1$~(K)   \\ \hline
		SiO$_2$ & 384   &0.0  & 17.96                       & 2.20         & 3600 & 2500 \\
		NS5    & 414   &16.7   & 18.07                      & 2.35         & 3000 & 2000 \\
		NS3    & 396   &25.0  & 17.62                       & 2.43         & 2200 & 1500 \\ \hline
	\end{tabular}
	\caption{
		\label{tab: ab simu-parameters} 
		Simulation parameters. See the main text for the definitions of $T_0$ and $T_1$.}
\end{table*}

In order to generate samples that have surfaces, we cleaved the bulk liquid
samples along the $z$-axis and inserted a vacuum layer between the two
surfaces, thus creating a sample with a slab geometry. The height of
this vacuum layer was 18\AA, large enough to prevent the two surfaces
to interact with each other. Due to the cleavage process the structure
close to the surfaces was strongly out of equilibrium and hence we
re-equilibrated the sample at a temperature $T_0$
(see Tab~\ref{tab: ab simu-parameters}). Note that the presence of the free
surfaces requires that this equilibration is done with some caution:
On one hand, the temperature should be high enough to allow the atoms to
diffuse within a reasonable amount of time. On the other hand a temperature
that is too high will result in the evaporation of the surface atoms
and/or a large expansion of the sample. For this reason, the equilibration
temperature $T_0$ for silica and NS5 sample was identical to the one
at which the liquid was equilibrated, since the evaporation rate is
small, while for the Na-rich composition NS3, for which the rate is high, 
we had to choose a lower temperature, namely $2200$~K.

The time for equilibration at $T_0$ was around 12~ps, which is long
enough for the structure to relax. The samples were subsequently quenched
down to an intermediate temperature $T_1$ using a nominal cooling rate
of $5\times10^{14}$~K$/$s, and then to 300~K using a faster cooling rate
of $2\times10^{15}$~K$/$s. The temperature $T_1$ was 2500~K, 2000~K, and
1500~K for silica, NS5, and NS3 respectively, values that were chosen
such they are  below the glass transition temperature $T_g$ of
the simulated glasses which, due to the fast cooling rates, are above
the experimental $T_g$'s. Finally, the samples were annealed at room
temperature for another 3~ps.  All simulations were carried out using
the $NVT$ ensemble. For the calculation and analysis of the observables
of interest, we discarded the first 4~ps from the total length of the
runs at  $T_0$, and  0.5~ps at 300~K.
Finally the samples
were quenched to 0~K and relaxed, and then we calculated the dynamical matrix and
the Born charge tensors in order to compute the vibrational density of
states (VDOS) as well as the imaginary part of the dielectric function.
(See Ref.~~\cite{pedesseau_first-principles_2015-1} for
details).

The AIMD simulations were performed by using the Vienna \textit{ab initio}
package (VASP)~\cite{kresse_efficiency_1996,kresse_efficient_1996} which
implements the Kohn-Sham (KS) formulation of the density functional theory
(DFT)~\cite{kohn_self-consistent_1965,martin_electronic_2004} to
compute the electronic structure. For the exchange and correlation term,
we used  the generalized gradient approximation (GGA) and the PBEsol
functional, respectively,~\cite{perdew_generalized_1996,perdew_restoring_2008}. The KS
orbitals were expanded in a plane-wave basis at the $\Gamma$ point and the
electron-ion interaction was described within the projector-augmented-wave
formalism~\cite{blochl_projector_1994,kresse_ultrasoft_1999}. The
plane-wave basis set included all components with energies up to 600~eV. For
solving the KS equations, the residual minimization method-direct
inversion was used in the iterative space, and the electronic convergence
criterion was fixed at $ 1\times 10^{-6}$~eV during the glass production
process and at $5\times 10^{-7}$~eV for the geometric optimization
procedure.

The time step for the simulations was 1~fs and temperature was
controlled by a Nos\'e thermostat~\cite{nose_molecular_1984}. We
note that the simulation parameters chosen here are similar
to the ones of previous \textit{ab initio} studies of silicate liquids and
glasses in the bulk~\cite{pedesseau_first-principles_2015,pedesseau_first-principles_2015-1,sundararaman_new_2018,sundararaman_new_2019},
and which have demonstrated that the resulting properties of the liquid
and glass compare very well with experimental results.

\begin{figure*}[t]
\centering
\includegraphics[width=0.3\textwidth]{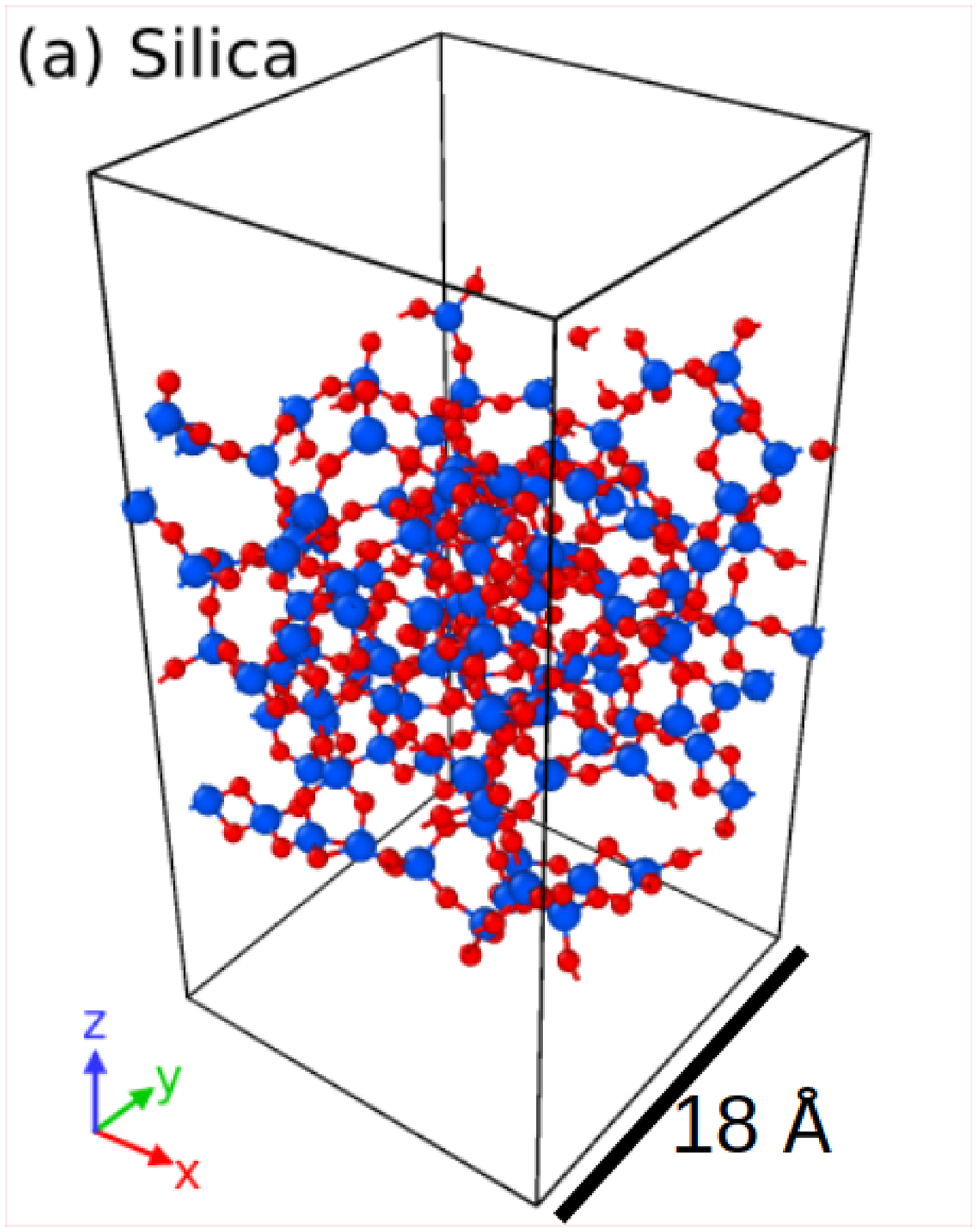}
\includegraphics[width=0.3\textwidth]{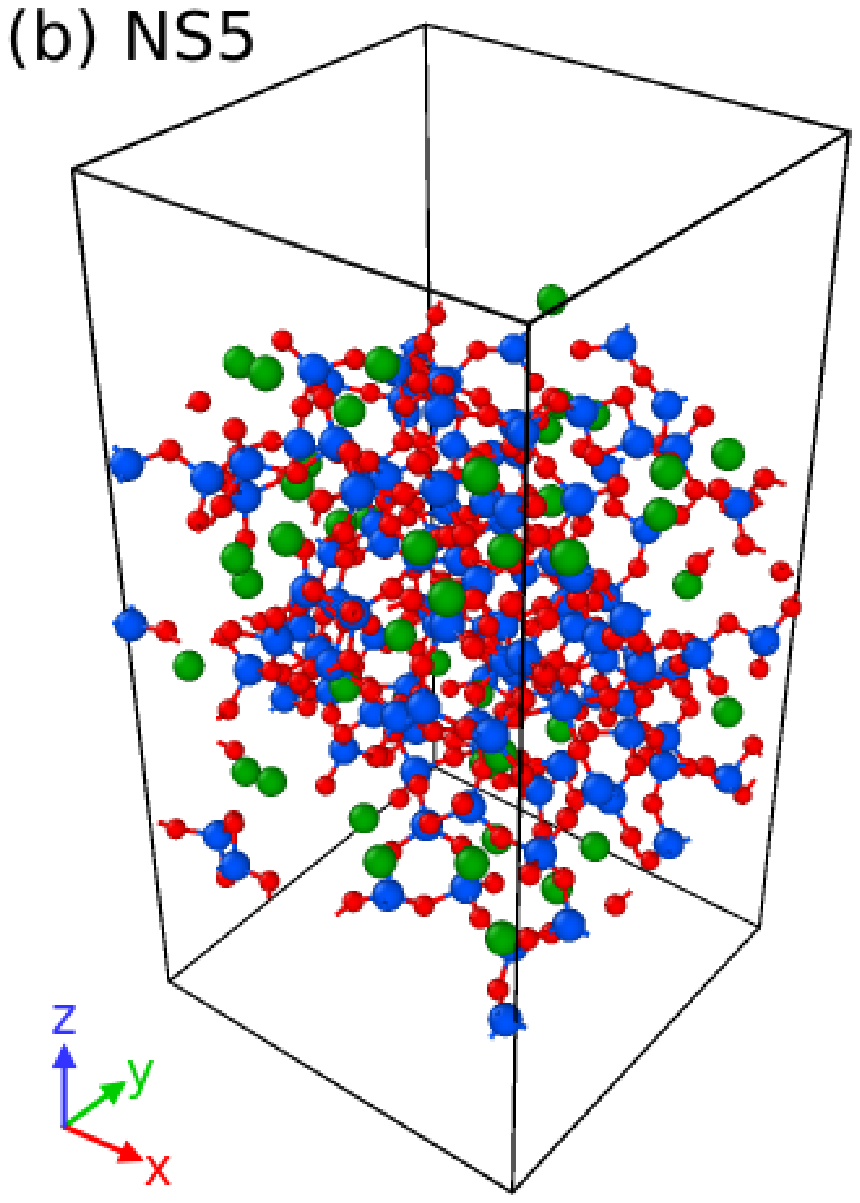}
\includegraphics[width=0.3\textwidth]{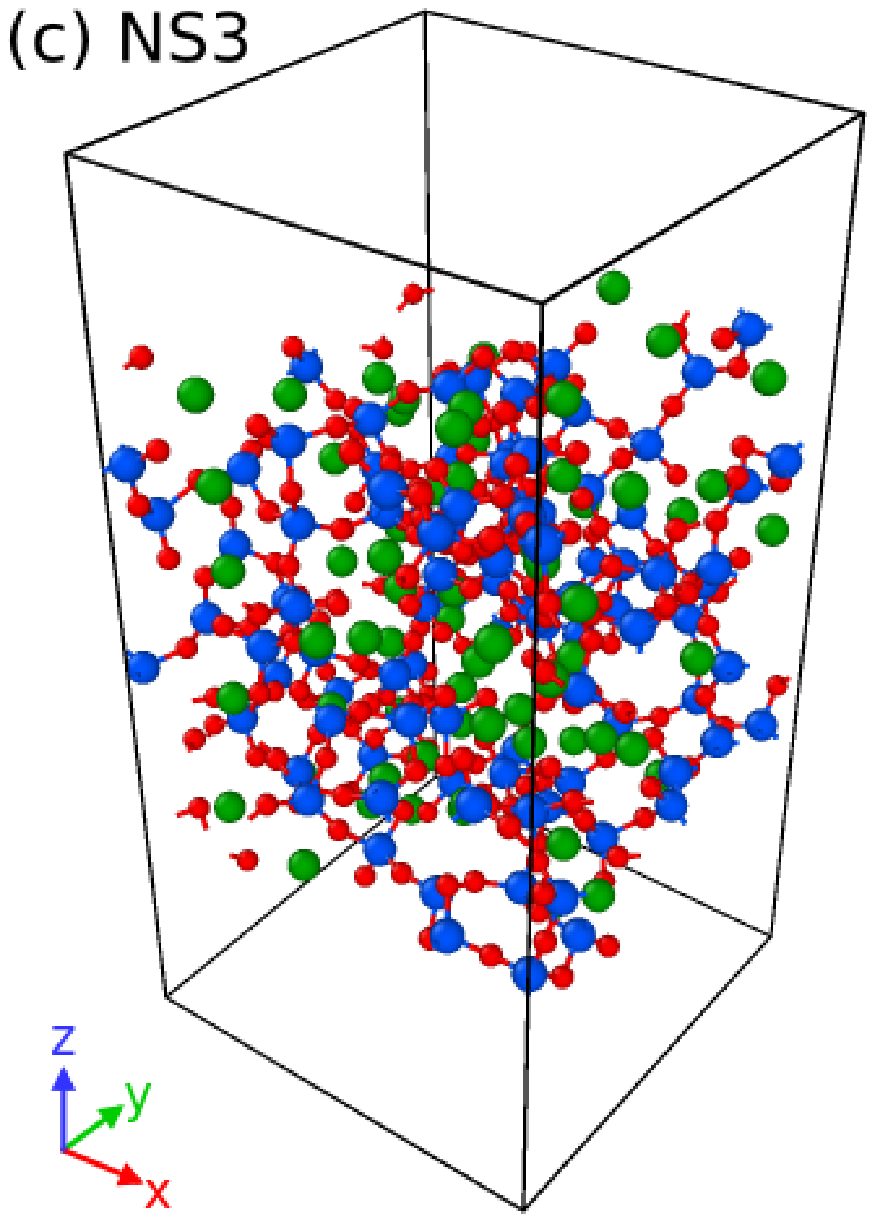}
\caption{Snapshots of the atomic structure of the three glasses
at 300~K. Si, O, and Na atoms are represented by spheres in blue, red, and green,
respectively. The sticks represent Si-O bonds with bond length smaller
than 2~\AA. 
}
\label{fig:ab nsx-snapshots-300K}
\end{figure*}

\section{Structure}\label{sec:structure}

In this section we describe how to identify the surface and interior
domains of our sandwich samples. Subsequently we will characterize their atomic
structure in terms of pair and bond angle distribution functions as well
as the concentration of the various species and local environments. These features
will then be discussed with respect to both their compositional dependence
and  location in the two sub-domains, i.e. surface and interior.

\begin{figure*}[th]
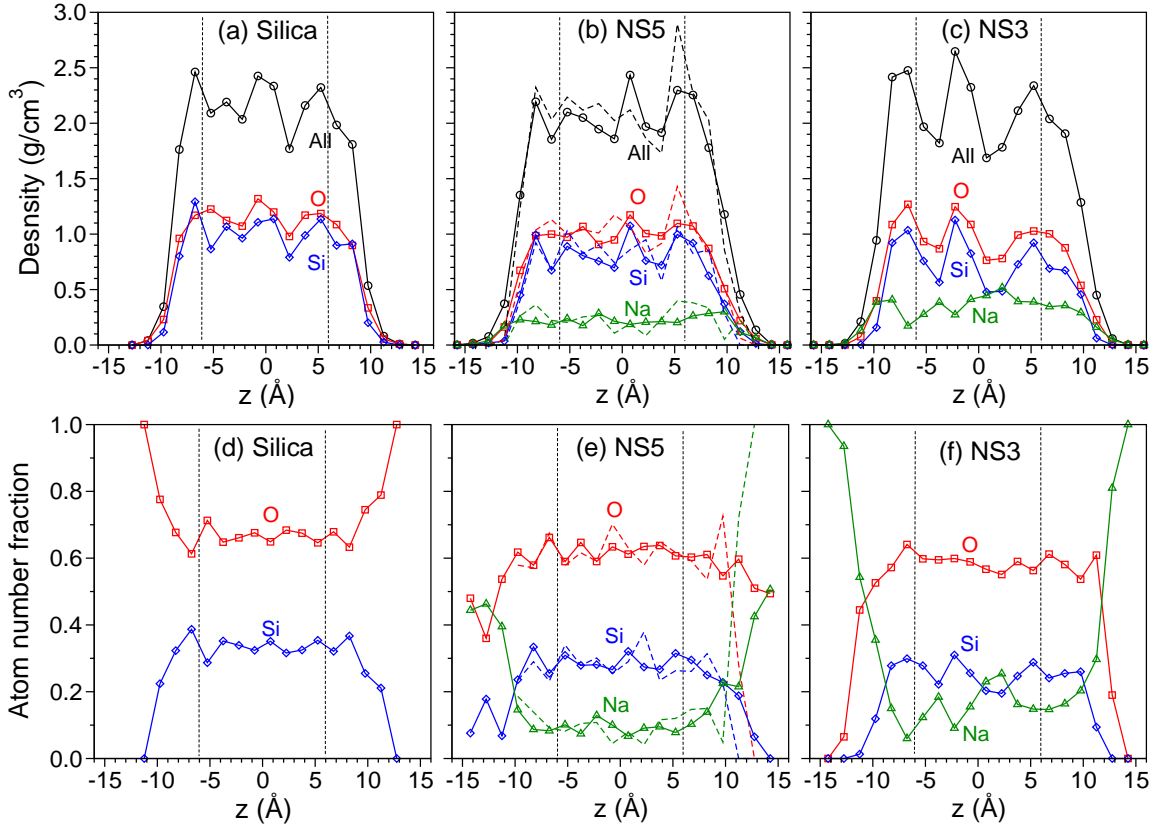

\centering
\includegraphics[width=0.94\textwidth]{fig2abc-ab-nsx-r2-massdensity.eps}
\includegraphics[width=0.94\textwidth]{fig2def-ab-nsx-r2-numfraction.eps}
\caption{Atomic distribution along the $z-$direction. Panels (a)-(c) are
the mass density profiles for silica, NS5, and NS3. Panels (d)-(f) are the
atomic number fraction along the $z-$direction for silica, NS5, and NS3. In
all graphs, the solid lines with symbols are for the liquids at temperature $T_0$,
see Table~\ref{tab: ab simu-parameters}. The dashed lines are the corresponding quantities 
for glasses at 300~K and for clarity are shown only for NS5. The vertical dashed
lines indicate the boundary between the surface and interior layers. }
\label{fig:ab nsx-density-numfrac}
\end{figure*}

\subsection{Defining the surface domain}
Figure~\ref{fig:ab nsx-snapshots-300K} shows snapshots of the simulation
boxes of the glasses for the three compositions at 300~K. One sees
that the slab has an atomic disordered network structure which
becomes increasingly depolymerized with the addition of Na$_2$O. In
Fig.~\ref{fig:ab nsx-density-numfrac} we plot the density and atomic
concentrations for the liquid state ($T=T_0$) as a function of the $z-$coordinate,
i.e.~perpendicular to the surface.  (Note that in the $z$-direction
the center of mass of the sample is defined to be at $z=0$.)  For all
three compositions the total density distributions show a relatively
flat region for $|z|\leq 6$ \AA, with densities around 2.2 g$/$cm$^3$
(silica), 2.3 g$/$cm$^3$ (NS5) and 2.4 g$/$cm$^3$ (NS3), Fig.~\ref{fig:ab
nsx-density-numfrac}(a)-(c). However, as we will discuss below, although
these values are similar to the  bulk densities, reported in
Tab.~\ref{tab: ab simu-parameters}, this similarity does not imply that
the inner region of the sandwich sample presents the same properties as
a real bulk glass.

Also included in Fig.~\ref{fig:ab nsx-density-numfrac}(b) are the
density distributions for the NS5 glass at 300~K (dashed lines). Within
the available statistics we do not find significant differences between
the distributions for the liquid and the ones for the glass, except for
the fact that the latters are slightly narrower due to  the shrinking of
the sample during the cooling process, resulting in a density of the
interior part which is slightly higher than that of the liquid. These
observations hold for all three compositions. This fact allows us to use
in the following a simple criterion for defining the different domains for
both liquids and glasses: Atoms having a $z$-coordinate with $ |z|\leq
6$ \AA\ will be defined to belong to the interior part of the sample,
while atoms with a $z$-coordinate  beyond this threshold are defined to belong to
the surface  layers. A similar strategy for defining surfaces was also
used in previous simulation studies of glass surfaces, see for examples
Refs.~\cite{roder_structure_2001,rarivomanantsoa_classical_2001,ren_surface_2017,halbert2018modelling}.

Figures~\ref{fig:ab nsx-density-numfrac}(d)-(f) depict the profiles
of the atomic number fraction along the $z-$direction. For silica,
we find that the concentration of oxygens in the surface regions
is higher than in the interior layer, indicating that the top
2~\AA~of the surface layers are enriched in O, fall slighty
below the bulk value at around 3-4~\AA, and, after a small
secondary peak, attains the bulk value, observations that are in
agreement with previous classical and ab initio simulations of silica
surface~\cite{garofalini_molecular_1983,roder_structure_2001,rimola_silica_2013,halbert2018modelling}.
For the sodium silicate glasses, i.e., NS5 and NS3,  the surface layers
are strongly enriched in Na and consequently the fractions of Si and O
decrease. This Na enrichment reaches about a factor of 3 (5) with respect
to the bulk value of NS5 and NS3, respectively. For the NS3 surface the
Na fraction reaches in fact 100\%, i.e.~the whole outermost layer is composed
by pure Na. These findings are consistent with experimental observations of
the surfaces of alkali silicate glasses by using LEIS
spectroscopy~\cite{kelso_comparison_1983,almeida_low-energy_2014,almeida_low-energy_2016}
as well as with recent findings from  classical molecular
simulations of sodosilicate glasses with reactive force
fields~\cite{mahadevan2020hydration}.

\begin{figure}[ht]
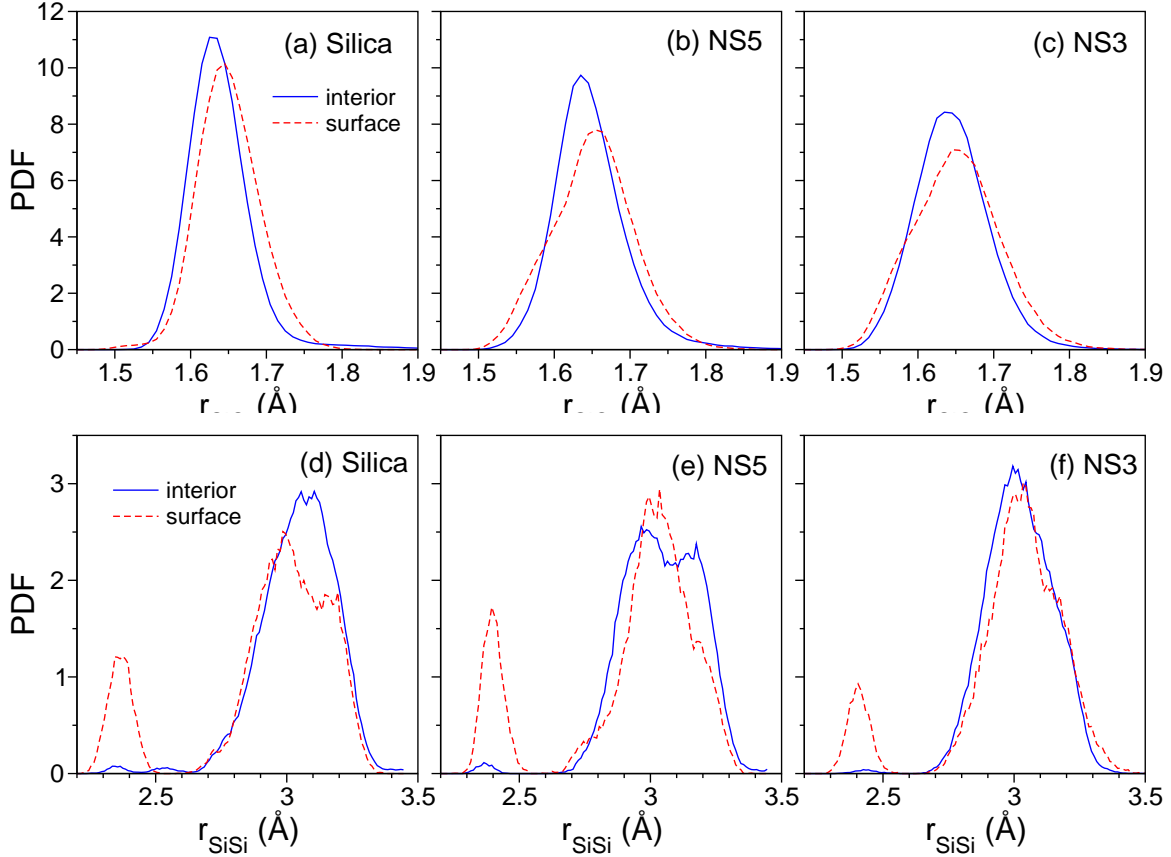

\centering
\includegraphics[width=0.95\textwidth]{fig3abc-ab-nsx-r2-pdf-sio.eps}
\includegraphics[width=0.95\textwidth]{fig3def-ab-nsx-r2-pdf-sisi.eps}
\caption{Probability distribution function  of nearest neighbor distance 
for the glasses at 300~K. Upper and lower panels are for Si-O and Si-Si
pairs, respectively. From left to right the compositions are silica,
NS5 and NS3, the solid  lines being for the surface domains, while the
dashed lines for the interior ones.
}
\label{fig:ab nsx-gr-sio-sisi}
\end{figure}

\subsection{Bond lengths and angles}
Further insight into the structure of the glass surfaces can be obtained
by investigating the interatomic distances, bond angles and by the
identification of the main local structural motifs. To start, we show in
Fig.~\ref{fig:ab nsx-gr-sio-sisi} the (normalized) probability distribution function
(PDF)  of the nearest neighbor distances for the Si-O and Si-Si pairs
calculated for the interior and surface regions. For the Si-O pair,
panels (a)-(c) in Fig.~\ref{fig:ab nsx-gr-sio-sisi}, we clearly see
that the distribution for the surfaces shifts to larger distances with
respect to the interior ones. In addition, the Si-O PDFs are broader
for the surface domains of NS5 and NS3 systems, which reflects the
increase of the network depolymerization and disorder with respect to
the corresponding properties in the interior. 
As we will see below, in the surface domain the fraction
of non-bridging oxygens (NBO) is indeed enhanced with respect to the
interior, while the fraction of bridging oxygens (BO) is smaller.
(Note that the BO atoms are oxygen atoms bonded to two silicons, while
NBOs connect to only one silicon). For the Si-Si pair, Fig.~\ref{fig:ab
nsx-gr-sio-sisi}(d)-(f), a prominent feature is the peak at around
2.4~\AA, in particular for the surfaces. This length corresponds
to the Si-Si first neighbor distances between two tetrahedra that share
an edge and thus form a two-membered (2M) ring, a structural
defect often found  on  dehydroxylated and dry surface of silica
glass~\cite{morrow1976infrared,michalske1984slow,bunker1989infrared,dubois1993bonding,dubois1993reaction,grabbe1995strained,tielens2019Characterization}.
Hence our results show that such structures are not only present at the
surface of silica but also in the sodo-silicate systems.

\begin{table}[htb]
\small
\center
\begin{tabularx}{13cm}{lYYY}
\hline
Glass         & Silica        & NS5           & NS3           \\ \hline
Bond & int.~[\AA] / surf.~[\AA]  & int.~[\AA] / surf.~[\AA]  & int.~[\AA] / surf.~[\AA]  \\ \hline
Si-Si          & 3.041 / 2.928 & 3.041 / 2.927 & 3.021 / 2.986 \\
csSi-csSi       &3.048 / 3.006  &3.041 / 3.059  &3.023 / 3.024 \\
csSi-esSi       &3.055 / 3.058  &3.189 / 3.000  &2.956 / 3.108 \\
esSi-esSi       &2.443 / 2.483  &2.370 / 2.408  &2.437 / 2.408 \\ \hline
Si-O          & 1.638 / 1.650 & 1.648 / 1.651 & 1.645 / 1.651 \\
Si-NBO        & - / 1.563     & 1.578 / 1.576 & 1.588 / 1.585 \\
Si-BO         & 1.638 / 1.652 & 1.652 / 1.664 & 1.655 / 1.670 \\
esSi-esBO       & 1.721 / 1.684 & 1.678 / 1.695 & 1.704 / 1.695 \\  \hline
Na-O       & - & 2.423 / 2.362&    2.423 / 2.375 \\
Na-NBO  & - & 2.276 / 2.264&    2.347 / 2.315 \\
Na-BO   & - & 2.526 / 2.531&    2.545 / 2.494 \\ \hline
~\end{tabularx}
\caption{
\label{tab:ab-bond-liquid-glass}
Average bond lengths for the sandwich glass samples at 300~K for both
surface and interior domains. 
csSi and esSi denote, respectively, corner-sharing and edge-sharing Si. 
}
\end{table}

In Tab.~\ref{tab:ab-bond-liquid-glass}, we report the average first
neighbors distances for the Si-O, Si-Si and Na-O pairs, as well
as the ones related to 2M-rings. The bond lengths with respect to
oxygens are further decomposed with respect to the two species BO and
NBO. For both domains, we see that the distances Si-NBO and Na-NBO
are significantly shorter than the Si-BO and Na-BO distances, as already pointed
out in simulations for bulk systems and in agreement with experimental
findings~\cite{ispas_structural_2001,tilocca2006structural,angeli_insight_2011,pedesseau_first-principles_2015-1,kilymis_vibrational_2019}.
For the sodo-silicates we find that the average Na-O distances in the
surface domains are shorter than the ones in the interior as a consequence
of the Na enrichment of the surface domains (see Fig.~\ref{fig:ab
nsx-density-numfrac}, e-f) which leads to the increased fraction of
NBOs and thus making the network less polymerized and hence with less
constraints compared to the interior part. The same trend has also been
found in a recent study of the surface structure of sodium silicates
glasses using classical MD~\cite{zhang_surfcl_2020}.

In silicate glasses, the Si-Si first neighbor distance is a measure of
the inter-tetrahedral distance between two corner sharing (cs) tetrahedra,
with typical values around $3.00-3.08$~\AA~\cite{wright1994neutron}. This
range is compatible with the values we find in the interior domain of our
three glasses, while for the surface domains this distance is shorter by 2-3\%,
see Tab.~\ref{tab:ab-bond-liquid-glass}. A further decomposition of
the structure into local motifs shows that this reduction in the Si-Si
distance for the surfaces is due to the presence of edge-sharing (es)
tetrahedra forming the 2M rings mentioned above. The 2M rings found in
our samples have tetrahedra that are strained, characterized by short
Si-Si distances, elongated Si-O bonds and reduced Si-O-Si and OSiO
angles (see below). As a consequence the esSi-esSi distance gives rise to an
additional peak seen in Fig.~\ref{fig:ab nsx-gr-sio-sisi}(d)-(f) 
located somewhat around 2.4~\AA (see also Tab.~\ref{tab:ab-bond-liquid-glass}).

Table~\ref{tab:ab-bond-liquid-glass} shows that in our
sandwich samples the esSi-esSi distance is close to 2.4~\AA\,  and  within the accuracy of our data, independent of the composition and whether the atoms
are located in the surface or in the interior domain, values that
compare well with results obtained from previous classical MD simulations
\cite{garofalini_molecular_1983,feuston_topological_1989,roder_structure_2001,rarivomanantsoa_classical_2001,du_molecular_2005,halbert2018modelling},
showing that this distance is not very dependent on the
potential used for the simulations.  The 2M rings found in
our samples are also characterized by Si-O bonds that are
stretched with respect to those in  corner-sharing tetrahedra,
and the values reported in Tab.~\ref{tab:ab-bond-liquid-glass}
are in good agreement with those found in classical MD
simulations~\cite{garofalini_molecular_1983,feuston_topological_1989,du_molecular_2005}
as well as in a recent AIMD investigation of dehydroxylated silica
surfaces~\cite{comas2016amorphous}. This elongation of the bond is
also compatible with DFT calculations of cristalline fibrous silica
containing chains of 2M rings~\cite{hamann1997energies} as well
as Hartree-Fock calculations of clusters and molecules containing
2M rings~\cite{okeeffe1984defects,bunker1989infrared} which found
esSi-esBO bond lengths around 1.67~\AA~and esSi-esSi distances between
2.38-2.42~\AA. Finally we mention that the presence of elongated esSi-esBO
bonds together with small Si-O-Si angles as structural fingerprints
of 2M rings is also consistent with the findings from a recent DFT study considering
the surfaces of $\beta$-cristobalite~\cite{le2018structural}. As a consequence we
conclude that i) for silica the geometrical properties of our 2M rings are compatible
with previous results and ii) the geometry of these rings are basically independent
of the environment of the ring.

\begin{table}[htb]
        \small
        \center
        \begin{tabularx}{\textwidth}{lYYYYYY}
                \hline
                & \multicolumn{2}{c}{Silica}    & \multicolumn{2}{c}{NS5}       & \multicolumn{2}{c}{NS3}       \\ \cline{2-7}
                & liquid        & glass         & liquid        & glass         & liquid        & glass         \\
                \%          & int. / surf.  & int. / surf.  & int. / surf.  & int. / surf.  & int. / surf.  & int. / surf.  \\ \hline
                $N_{\rm domain}$          & 65.8 / 34.2   & 66.7 / 33.3   & 58.6 / 41.4   & 60.7 / 39.3   & 58.5 / 41.5   & 62.9 / 37.1   \\ \hline
                Si          & 33.2 / 33.7   & 33.3 / 33.4   & 29.1 / 26.0     & 29.1 / 25.7   & 25.5 / 24.4   & 25.8 / 23.7   \\
                O           & 66.8 / 66.3   & 66.7 / 66.6   & 61.9 / 60.0     & 62 / 59.7     & 58.3 / 58.4   & 58.3 / 58.4   \\
                Na          & 0 / 0         & 0 / 0         & 9.1 / 14.0      & 8.9 / 14.6    & 16.3 / 17.2   & 15.9 / 17.9   \\ \hline
                Si$^3$         & 2.3 / 3.7     & 0 / 1.6       & 0.7 / 1.7     & 0 / 0         & 0.1 / 0.8     & 0 / 0         \\
                Si$^4$         & 29.6 / 28.8   & 32.9 / 31.8   & 27.2 / 23.6   & 27.5 / 25.7   & 24.3 / 23.3   & 25.4 / 23.7   \\
                Si$^5$         & 1.2 / 0.8     & 0.4 / 0       & 1.2 / 0.6     & 1.6 / 0       & 1.1 / 0.3     & 0.4 / 0       \\ \hline
                NBO         & 2.4 / 4.6     & 0 / 2.3       & 8.1 / 16.7    & 6.6 / 15.6    & 14.3 / 19.7   & 14.9 / 19     \\
                BO          & 64.4 / 61.7   & 66.7 / 64.3   & 53.8 / 43.3   & 55.4 / 44.1   & 43.9 / 38.7   & 43.5 / 39.3   \\  \hline
                esBO        & 4.9 / 11.1    & 1.4 / 12.8    & 3.4 / 7.5     & 1.6 / 9.8     & 1.8 / 4       & 0.8 / 4.1     \\
                esSi        & 4.6 / 11.3    & 1.6 / 12.5    & 3.3 / 7.1     & 1.2 / 10.5    & 1.6 / 4.1     & 0.4 / 4.8     \\  \hline

\end{tabularx}
\caption{
\label{tab:ab struc-liquid-glass}
Percentages of various atomic species in the interior and surface domains
for the silica and sodo-silicate samples. Liquids correspond to simulation
at $T_0$ (see Table~\ref{tab: ab simu-parameters}), and glasses  are at
300~K. On the first row, $N_{\rm domain}$  denotes the percentage of atoms
in a specific domain with respect to   the total number of atoms of the
sample. The proportions of the atomic species are given relative to their
concentration in the considered domain. Note that, for the surface
domain, we give the sum of the amounts on the top and bottom surface layer.
  }
\end{table}

In order to characterize the structure of our sandwich systems in a more
quantitative manner we have determined the fractions of various atomic
species present in the interior and surface domains, and the data are
summarized in Table~\ref{tab:ab struc-liquid-glass}.  One recognizes
that for the sodo-silicate systems the surface domains are enriched in sodium,
in agreement with the atomic distribution along the $z-$
axis, shown in Figs.~\ref{fig:ab nsx-density-numfrac}(d)-(f). For NS5
this enrichment is about 50\% while for NS3 it is still around 10\%,
independent whether one considers the liquid or the glass state.
Furthermore, we have decomposed in both domains the concentration
of the silicon and oxygen atoms with respect to  their coordination
numbers. (For this we used a cutoff distance of 2.0~\AA~to define bonded pairs.)  We find
that most Si atoms are 4-fold coordinated but also note the
presence of under-and over-coordinated atoms, Si$^3$ and Si$^5$,
respectively. For the systems with sodium we see that in the liquid
state there is a small concentration of 5-fold coordinated Si for
both surface and interior domains but that during the quench these
defects disappears in the surface domains, while a very small number
are still present in the interior domains, possibly as a consequence of
the high quench rate~\cite{pedesseau_first-principles_2015}. For the
3-fold Si we note that in the liquid state they are more concentrated
in the surface layers than in the interiors and that in the glassy state
they are absent for the sodium silicate systems

From the data reported in Tab.~\ref{tab:ab struc-liquid-glass} one also recognizes that the surface domains have a
significantly higher concentration of NBOs than the interior. With
increasing Na$_2$O content, the concentration of surface NBO increases,
and this may account for the reduction of both under- and over-coordinated
silicons since with increasing Na content more NBO are formed, allowing
the Si atoms to grab/shed O atoms and hence to form a more  regular local
environment, i.e.~becoming four-fold coordinated.

The last two rows in Tab.~\ref{tab:ab struc-liquid-glass} give the percentages of the silicon and
oxygen atoms that form 2M rings (labelled esSi and esBO respectively). As
expected, these rings are more abundant on the surface than in the
interior, by a factor of around 2 in the liquid state and a factor 5-8
in the glass samples. This temperature dependence is mainly due to the
$T$-dependence of the concentration of the 2M rings in the interior since
on the surface this concentration is basically independent of $T$. This indicates
that  these structures are energetically very
unfavorable in the bulk  while   they are a energetically reasonable
building block in the presence of a free surface.

From the numbers of 2M rings we can calculate their area density by
dividing this number by the surface area. For silica,  we find a density of 1.5/nm$^2$ which has to be compared with the estimate obtained
from IR experiments which give values of 0.2 to 0.4/nm$^2$. 
(As discussed
below, 2M rings have a spectroscopic signature in the absorption IR
spectrum, with two peaks at 888 and 908~cm$^{-1}$ and a weak shoulder at
932~cm$^{-1}$~\cite{morrow1976infrared,bunker1989infrared,ferrari1995reactions,grabbe1995strained,chiang1993first}.)
Also previous simulation studies on dry silica,
have reported smaller densities of
2M rings and this might be rationalized by the fact that most of
these simulations were carried out using classical MD approaches thus
allowing for quench rates that are considerably lower than the one used
in present study~\cite{rarivomanantsoa_classical_2001,halbert2018modelling}. To our knowledge, the models of silica surface labelled
as \textit{ab initio} in the literature were initially prepared by melting
and quenching a liquid silica using effective, i.e.~classical, potentials
and the obtained structure was processed within a first-principles
framework only at 300~K (see for example Ref.~\cite{rimola_silica_2013}
and references therein). The present samples are hence the first ones
generated by the quench of a liquid surface within an AIMD approach,
admittedly with a very high quench rate which prevents the annealing
and relaxation of the glass surface.

\begin{figure}[htb]
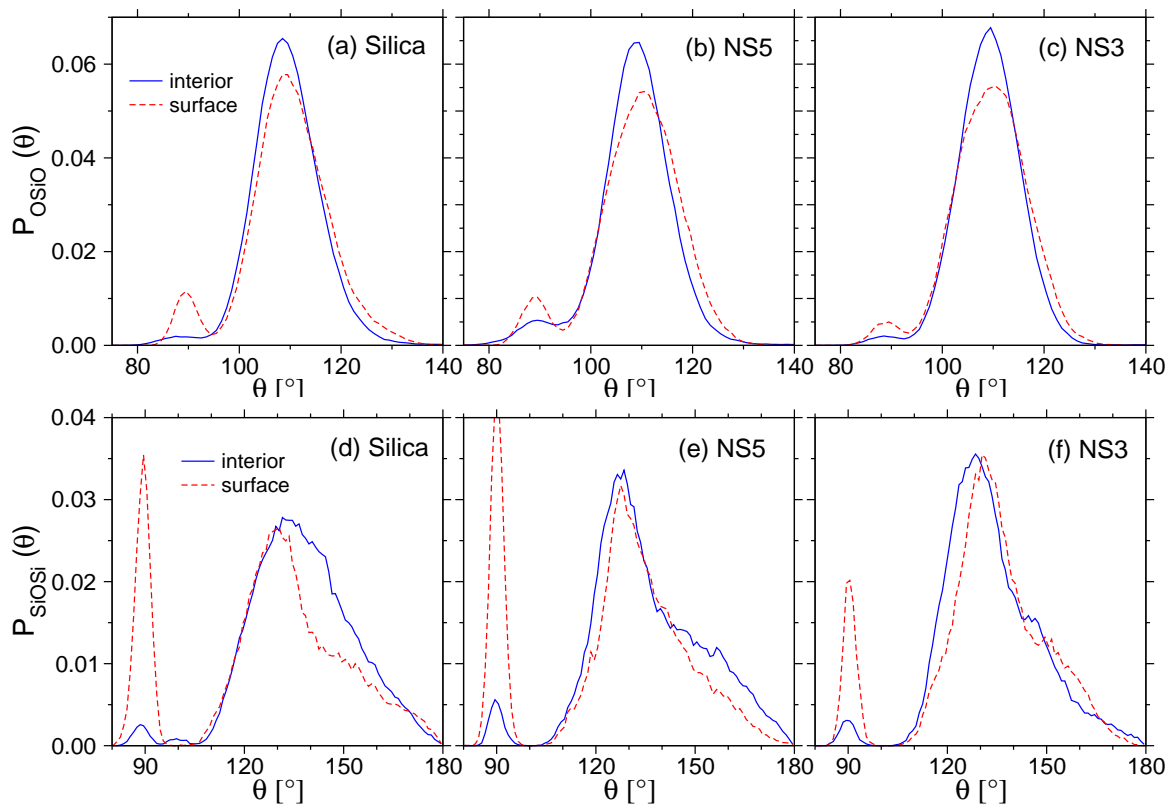

\centering
\includegraphics[width=0.95\textwidth]{fig4abc-ab-nsx-r2-bad-osio.eps}
\includegraphics[width=0.95\textwidth]{fig4def-ab-nsx-r2-bad-siosi.eps}
\caption{Bond angle distribution. Upper and lower panels are for O-Si-O
and Si-O-Si angles, respectively. From left to right the compositions are
silica, NS5, and NS3. 
}
\label{fig:ab nsx-bad-osio-siosi}
\end{figure}

More insight into the structural differences between the surface
and interior domains and the compositional effect can be obtained by
computing the bond angle distributions (BAD) shown in Fig.~\ref{fig:ab
nsx-bad-osio-siosi} for the glass samples. 
The main peak in the BAD for O-Si-O is located at around
109\degree, the expected angle for a perfect tetrahedron, panels
(a)-(c). One also notices that the distribution of surface O-Si-O
angle is slightly wider than the interior, which indicates that the
[SiO$_n$] units on the surface are more distorted than the interior
ones. For pure silica, we see that the main peak of the Si-O-Si BAD
($>100^\circ$), panel (d), is narrower and more asymmetric for the
surface domain. The peak at around 130$^\circ$ is similar to the one
found in the NS5 and NS3 systems, panels (e) and (f), i.e.~the glasses that
are more depolymerized. 
For the NS5 system, panel (e), the mentionned asymmetry is still present but
less pronounced, while for the Na-rich glass it has basically disappeared
due to the presence of Na atoms. For the interior domains, the main
peak becomes sharper and shifts to smaller angle with the addition of
Na$_2$O, a trend pointed out also in a previous \textit{ab initio} study
of bulk sodium silicate glasses~\cite{kilymis_vibrational_2019}.

For both the O-Si-O and Si-O-Si BADs, we observe a peak at
around 90\degree, which is due to the 2M rings. This peak is
more pronounced for the surfaces, which is consistent with the
structural data discussed above, i.e.~the presence of small
Si-Si distances and a significant fraction of the esSi and
esBO in the surface domains. The location of these peaks 
are in qualitative agreement with earlier MD simulations with classical 
potentials~\cite{garofalini_molecular_1983,feuston_topological_1989,rarivomanantsoa_classical_2001,du_molecular_2005,halbert2018modelling}
and also the aforementioned DFT~\cite{hamann1997energies} and molecular orbital
calculations~\cite{okeeffe1984defects,bunker1989infrared} obtained
optimized structures of edge-sharing tetrahedra  with similar highly
distorted Si-O-Si angles around  $90\degree$.

\begin{figure}[htb]
\centering
\includegraphics[width=0.98\textwidth]{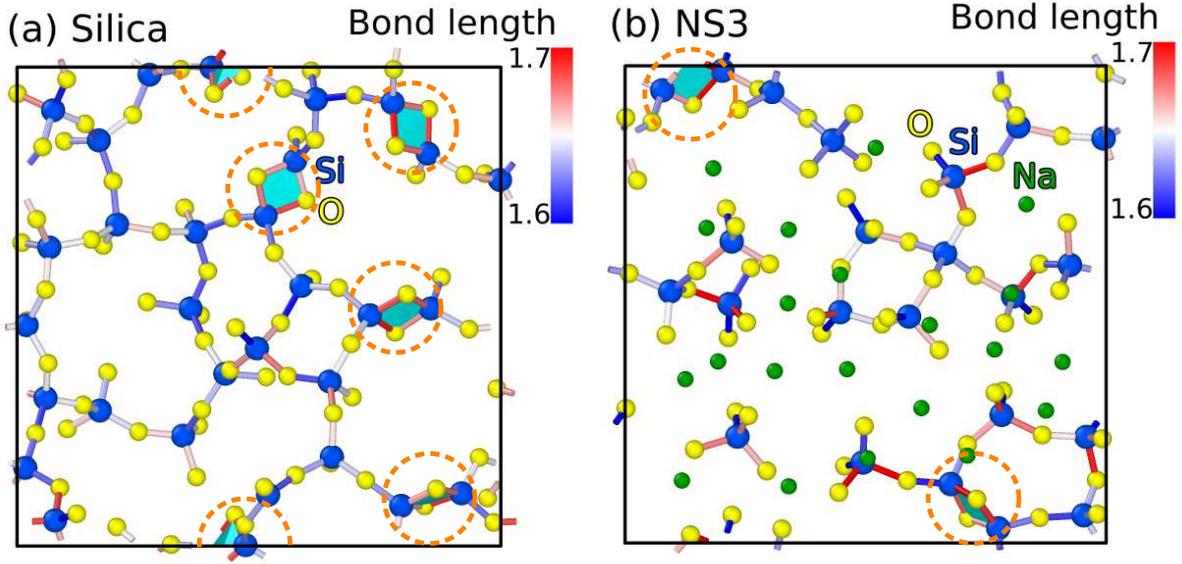}
\caption{Top view of snapshots showing  structural motifs at the surfaces
of (a) silica and (b) NS3 glasses.  Only the outermost layer of silicon
atoms and their nearest neighbors O and Na atoms are shown (see the text
for the construction method).  Two-membered rings  are highlighted with
a circle. Only Si-O bonds are shown and a color code is used taking into
account their length, given in \AA. The visualization was realized using Ovito~\cite{stukowski_visualization_2010}.
}
\label{fig:ab nsx-surf-snapshot}
\end{figure}

In order to visualize some of the above mentioned structural features
of the surfaces and their compositional differences, we show in
Fig.~\ref{fig:ab nsx-surf-snapshot} snapshots corresponding to the outermost
atoms of the surface domains of silica and NS3 glasses. In
order to select the atoms  shown in Fig.~\ref{fig:ab nsx-surf-snapshot},
we have first identified the Si atoms belonging to  the surface using
the tetrahedralization-based method proposed by Edelsbrunner and
M\"{u}cke~\cite{edelsbrunner_three-dimensional_1994}. The probing sphere
radius used for this algorithm was chosen as 3.2~\AA, i.e., around the nearest neighbor distances
of Si-Si (see Refs.~\cite{zhang_fracture_2020,zhang_surfcl_2020}
for details). The first nearest O and Na neighbors of these surface
silicons have then been found and included in the snapshots, whereas
the other atoms have been removed from the sake of clarity.  For silica,
Fig.~\ref{fig:ab nsx-surf-snapshot}(a), 2- to 9-M rings are found on the
surface. With the addition of sodium, the network of the atomic surface
layer of the NS3 glass, Fig.~\ref{fig:ab nsx-surf-snapshot}(b), becomes
less connected, and the proportion of 2M-rings is decreased: While for
silica we observe five 2M-rings, only two  are found for NS3. As argued
above, this difference in concentration is likely due to the fact that
2M-rings are strongly strained, with smaller Si-O-Si bond angles and
longer Si-O bond lengths with respect to the typical values (see bond
color code in Fig.~\ref{fig:ab nsx-surf-snapshot}). The presence of Na
can effectively relieve surface tension by breaking some of these Si-O
bonds (notably in small rings) and thus make the surface  energetically
more stable.

\section{Vibrational properties} \label{sec:vibrations}

\subsection{Vibrational density of states}
In this subsection we will discuss the vibrational properties of our
systems in terms of the total as well as partial vibrational density of
states (VDOS). 

After having relaxed to 0~K, we have determined and diagonalized
its dynamical matrix from which one can obtain the
total VDOS as

\begin{equation}
g(\omega)=\frac{1}{3N-3}\sum_{p=4}^{3N}\delta(\omega-\omega_p),
\label{eq1}
\end{equation}

\noindent
where $N$ is the total number of atoms in the sample, $\omega$ is the
frequency and $\omega_p$ is one of the 3$N$ eigenvalues of the dynamical
matrix. This total VDOS can be decomposed further into the contributions
from different species, allowing to define the partial VDOS

\begin{equation}
g_\alpha(\omega)=\frac{1}{3N-3}\sum_{p=4}^{3N}\sum_{I=1}^{N_\alpha}
\sum_{k=1}^{3}|{\bf e}_{I,k}(\omega_p)|^2\delta(\omega-\omega_p) \quad .
\label{eq2}
\end{equation}

\noindent
Here $\alpha\in \{ \mathrm{Si, O, Na, BO, NBO, csSi, esSi, csBO,
esBO}\}$, $N_\alpha$ is the number of particles of type $\alpha$, and
${\bf e}_{I,k}(\omega_p)$ is the part of the $3N$-component eigenvector
${\bf e}(\omega_p)$ that contains the three components of the particle
$I$. (Note that in Eqs.~(\ref{eq1}) and (\ref{eq2}) we do not consider the three trivial
translational modes of the system.) All vibrational spectra that will
be discussed in the following have been obtained by convoluting the
discrete distribution given by Eq.~(\ref{eq2}) with a Gaussian function
with a full width at half maximum of 30~cm$^{-1}$ and averaged over two
independent samples.

\begin{figure*}[ht]
\centering
\includegraphics[width=0.95\textwidth]{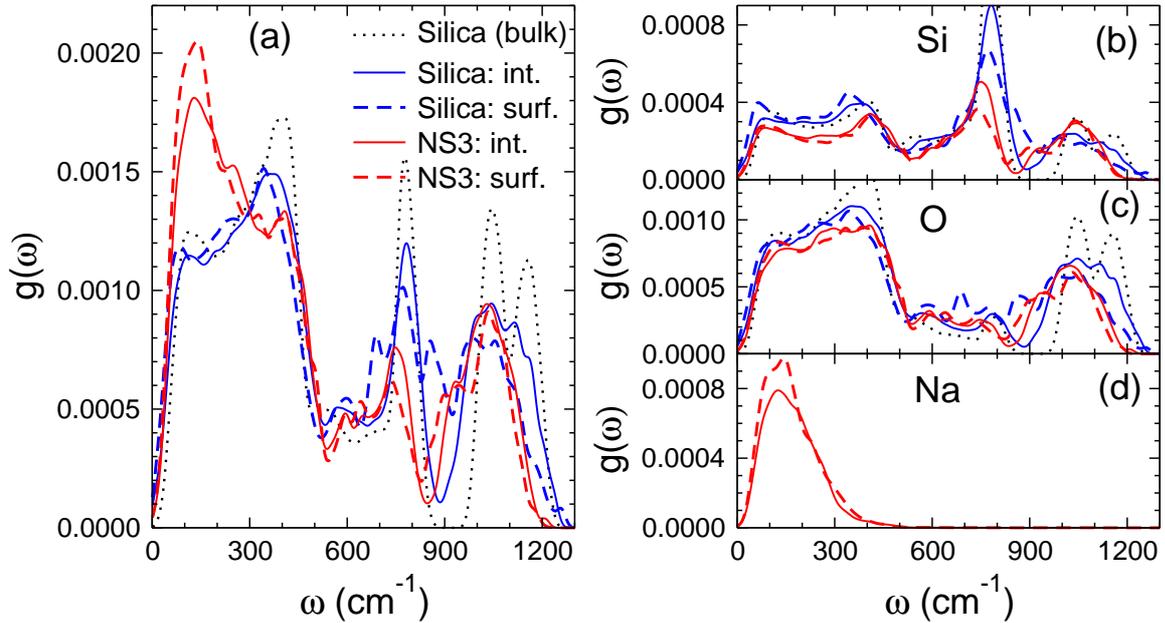}
\caption{(a) Total vibrational density of states (VDOS) of the
two sandwich glasses at 0~K for the interior and the surface domains. 
Panels (b)-(d) show the partial
VDOS for the Si, O and Na atoms, respectively. Also included in (a)
is the total VDOS for a bulk silica glass from \textit{ab initio}
calculations~\cite{sundararaman_new_2018}. The total VDOS in panel (a)
are normalized to unity, which is equal to the sum of the partials as
depicted in (b)-(d).
}

\label{fig:ab nsx-VDOS-tot-partial}
\end{figure*}

The total VDOS of the silica and NS3 systems are shown in Fig.~\ref{fig:ab
nsx-VDOS-tot-partial}(a), alongside the partial contributions
of their constituent atoms $\alpha \in \{{\mathrm {Si,\, O,\,
Na} }\}$, Fig.~\ref{fig:ab nsx-VDOS-tot-partial}(b)-(d). We
recognize that each of these distributions have three main bands:
A low-frequency band with $\omega<500$~cm$^{-1}$, a mid-frequency
band with $500<\omega<900$~cm$^{-1}$, and a high-frequency band
with $\omega>900$~cm$^{-1}$. In order to recognize the influence
of the surface on the spectra, we have included in Fig.~\ref{fig:ab
nsx-VDOS-tot-partial}(a) also  the total VDOS of a bulk silica glass sample
which was obtained from \textit{ab initio} calculations within a
framework that was similar to the used in the present work~\cite{sundararaman_new_2018}. The presence of the surfaces
makes that all the sharp peaks observed in the bulk glass at around 400,
800 and 1000~cm$^{-1}$ are significantly smeared out and that the high
frequency band is shifted to somewhat lower frequencies, making that
the gap between the mid and high frequency band is partially filled
up. In addition one recognizes that due to the surface the double
peak structure of the high frequency band is completely washed out.
The addition of Na$_2$O makes that the height of the peaks at 400 and
800~cm$^{-1}$ decreases further while the high frequency band is not
modified in a significant manner, although it does shift to lower $\omega$.
As it has been shown before, this softening is due to the depolymerization of
the network which increases the contribution from NBO-related
motions~\cite{zotov_calculation_1999,kilymis_vibrational_2019}.
Finally we note that the shape of the low frequency band
changes strongly in that a new peak at around 150~cm$^{-1}$ 
starts to grow with increasing Na concentration, a feature that is also
seen in spectra of bulk sodo-silicate glasses~\cite{kilymis_vibrational_2019}.

A better understanding of these changes can be obtained by
inspecting the partial VDOS, presented in Fig.~\ref{fig:ab
nsx-VDOS-tot-partial}(b)-(d). (Note that the sum of the
three partials gives the total VDOS shown in Fig.~\ref{fig:ab
nsx-VDOS-tot-partial}(a).) It is clearly seen from Fig.~\ref{fig:ab
nsx-VDOS-tot-partial}(b) that the band at around $800$~cm$^{-1}$ is
related to the vibrational motion of Si, in agreement with earlier
studies which have shown that the peak is related to the complex motion
of Si against BO~\cite{zotov_calculation_1999}. The decrease of the peak
height with the addition of Na can thus be expected to be related to the
(partial) breaking up of the network, i.e.~the decreasing number of BO.
Figure~\ref{fig:ab nsx-VDOS-tot-partial}(c) shows that oxygen is the
dominant contributor to the spectrum in the low frequency band and that
also in the high frequency band its partial VDOS is larger than the one
of Si.

\begin{figure*}[h]
\centering
\includegraphics[width=0.95\textwidth]{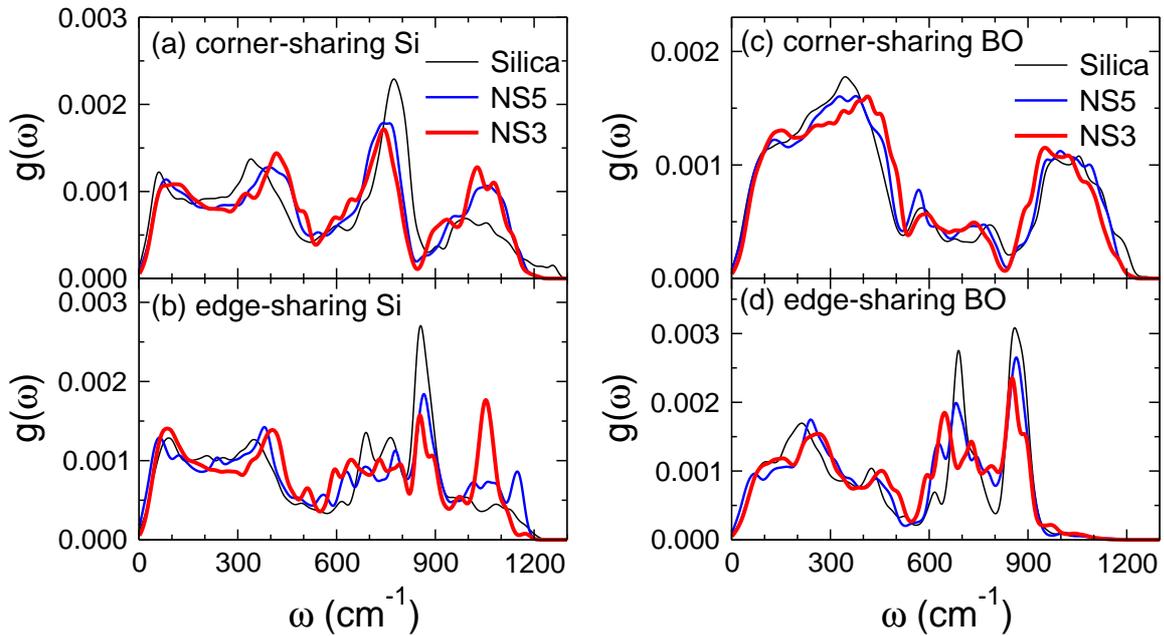}
\caption{Per-atom VDOS of the surface atoms. Panels (a) and (b) are for corner-sharing and edge-sharing Si atoms, respectively. Panels (c) and (d) are for corner-sharing and edge-sharing BO, respectively. All curves are normalized to unity.}
\label{fig:ab nsx-VDOS-es-cs}
\end{figure*}

The experimental IR spectra for silica surfaces show two strong
peaks at 888 and 908~cm$^{-1}$ and a shoulder at 932~cm$^{-1}$
~\cite{morrow1976infrared,bunker1989infrared,chiang1993first,grabbe1995strained,ferrari1995reactions}.
These features have been related to the presence of 2M rings, an assignment
which is supported by electronic structure calculations for small
terminated 2M ring clusters~\cite{bromley2003tworing} as well as of
dehydroxylated silica surface~\cite{ceresoli_two-membered_2000}. In
order to identify the vibrational signal of 2M rings in our sandwich
samples, we  have decomposed the partial VDOS of surface BO and Si into
contributions from edge-sharing and corner-sharing atoms, Fig.~\ref{fig:ab
nsx-VDOS-es-cs}. (We mention that to a first approximation the VDOS
of the corner-sharing atoms, Fig.~\ref{fig:ab nsx-VDOS-es-cs}(a) and
(c), are the same as the spectra for the bulk. In reality, however,
the presence of the surface gives rise to a slight modification of the
spectra.)  Panels~\ref{fig:ab nsx-VDOS-es-cs}(b) and (d) clearly show  that
esSi as well as esBO have a strong signal between 800 and 900~cm$^{-1}$,
a frequency range in which the spectra for csSi and csBO have low
intensity. The main peak in this range is at around 850~cm$^{-1}$, i.e.,~a
frequency which is somewhat lower than the experimental window which
ranges from 888 to 932~cm$^{-1}$, but a value that agrees well
with previous DFT calculations~\cite{ceresoli_two-membered_2000}.

In addition to the vibrational features discussed above, we note in
the partial VDOS for the esBO a further signature of the 2M rings in
the form of a pronounced peak at around 700~cm$^{-1}$, Fig.~\ref{fig:ab
nsx-VDOS-es-cs}(d).  This peak is completely absent in the spectra for the
csBO and its position shifts to lower frequencies with increasing Na$_2$O
concentration. At this frequency also the partial VDOS of the csSi shows a peak, but its intensity is not very high.
To the best of our knowledge, the existence of these peaks
for the vibrational spectra of 2M rings has not been reported before and
at present we do not know to which type of motion it corresponds to.

\begin{figure*}[t]
\centering
\includegraphics[width=0.95\textwidth]{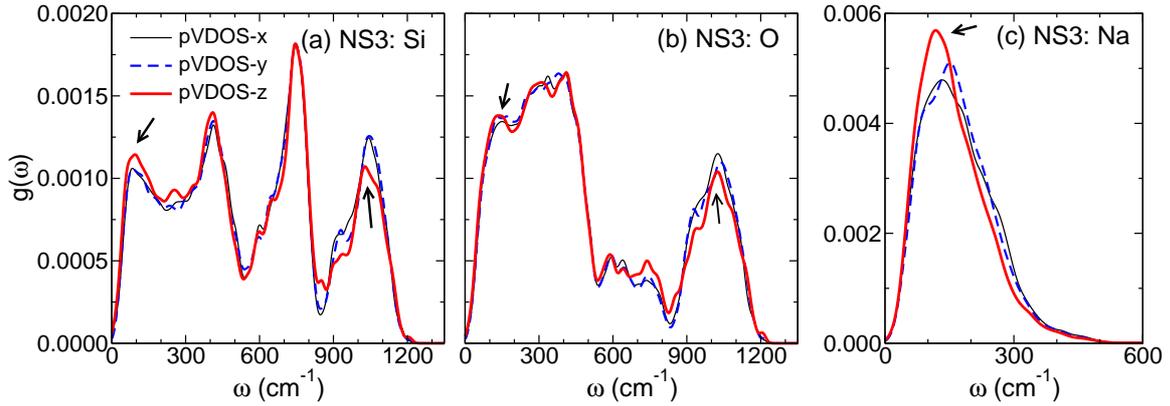}
\caption{Decomposition of the partial VDOS of NS3 into contributions from
different directions. Panels (a)-(c) are for Si, O
and Na, respectively. The arrows indicate the locations at which
the spectra depend significantly on the direction. All curves
are normalized to unity.\\
}
\label{fig:ab nsx-VDOS-ns3-xyz}

\end{figure*}

Due to the presence of the surface one can expect that the vibrational
modes are no longer isotropic and that hence also the VDOS will become
anisotropic.  That for the case of NS3 this is indeed the case is
demonstrated in Fig.~\ref{fig:ab nsx-VDOS-ns3-xyz} where we present the
partial VDOS as obtained for the three different directions: $x$ and $y$
parallel to the surface and $z$ orthogonal to it. We see that the curves
for the $x$ and $y$ directions coincide with high accuracy, indicating
that the error bars are small. The spectrum for the $z$-direction shows
significant deviations from the two other curves, see arrows, notably
at around $100$~cm~$^{-1}$, i.e.~the peak that is directly related
to the vibrational motion of the Na atoms. Panel (c) shows that the
vibrations in the $z$ direction are a bit softer than in the two other
directions (the peak is shifted to lower frequencies), a result that is
reasonable since the Na atoms are less constrained in the $z$
direction. The anisotropy can also be see in the high frequency band
in that the intensity of the spectrum in the $z$ direction for Si and O
is lower than the one in the orthogonal directions.  This result can be
rationalized by the fact that close to the surface the Si-O-network is  more anisotropic, since we have found, see Fig.~\ref{fig:ab nsx-density-numfrac},
that there is a layering effect in the composition.

\subsection{Infrared response}

\begin{figure*}[ht]
\centering
\includegraphics[width=0.95\textwidth]{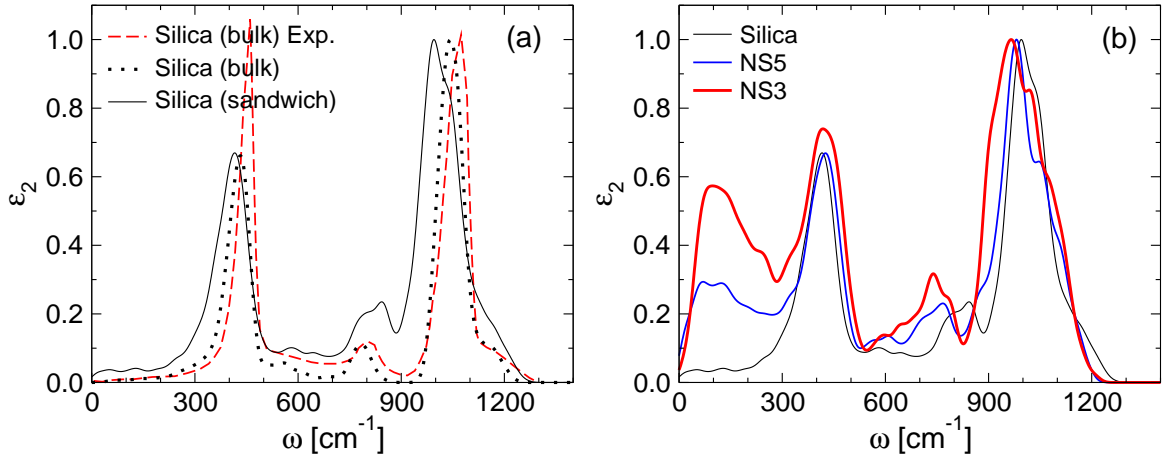}
\caption{Imaginary part of the dielectric function $\epsilon_2 (\omega)$
for bulk silica as well as for the three sandwich systems. Panel (a) shows
the calculated spectra of  $\epsilon_2 (\omega)$ for bulk (black dotted
line) and sandwich sample (black line), and the experimental spectrum
(red dashed line) for bulk amorphous silica~\cite{Philipp1998}. Panel
(b) shows the calculated  $\epsilon_2 (\omega)$  for the silica, NS5
and NS3 sandwich systems, black, blue and green full line, respectively.
}
\label{fig:ir-spectra}
\end{figure*}

In the previous subsection we have discussed the vibrational features
of our sandwich systems, focusing on the frequency and composition
dependences of the partial and total VDOS. In order to make a direct
connection to experimental data, it is useful to compute the IR response
of the samples. This quantity can be obtained directly from the frequency
dependence of the dielectric function $\epsilon(\omega)$ which can be
calculated from the vibrational eigenmodes and the Born effective charges
of the atoms. (The details of the method and the relevant relations
are documented in Ref.~\cite{pedesseau_first-principles_2015-1}).
In Fig.~\ref{fig:ir-spectra} we present $\epsilon_2(\omega)$, the
imaginary part of the dielectric function, for our three systems,
for bulk silica as well as the experimental spectrum from
Ref.~\cite{Philipp1998}. Since $\epsilon_2(\omega)$
has an \(\omega\)-dependence which is very similar to the one of the IR absorption, 
see Ref.~\cite{pedesseau_first-principles_2015-1}, we
present here the former quantity. We also recall that  the
IR experimental studies exhibiting the well-defined
frequency window between 888 and 932~cm$ ^{-1} $ assigned to 2M
rings~\cite{morrow1976infrared,bunker1989infrared,chiang1993first,ferrari1995reactions,grabbe1995strained}
are absorption spectra, thus motivating this choice.

Comparing in Fig.~\ref{fig:ir-spectra}(a) the theoretical spectra for the
bulk with the experimental data, we see that the simulation reproduces
correctly the three main resonances, although the peak positions are
down-shifted by about 25-30~cm$^{-1}$ and the height of the peak at
$\approx$430~cm$^{-1}$ is lower. These discrepancies might be due to the
small size of our system or related to the fact that the DFT functional
we have used in the present simulations is known to produce frequency
that are about 5\% too small~\cite{delapierre2016vibrational}.  However, in overall
the obtained agreement between the calculated $\epsilon_2(\omega)$ and
the experimental data is good and on this basis we can proceed to
understand the evolution of the IR response due to the presence of the
surface as well as to the composition.

Also included in panel~(a) is the spectrum for the silica sandwich
sample. We see that with respect to the corresponding bulk data the curve
is shifted to lower frequencies by about 30~cm$^{-1}$ and that the three
main peaks have become broader. These changes can be explained by the fact
that the defective structures present in the sandwich samples (NBO, 2M
rings) induce distortion of the glass network and this increased disorder
leads to a softening of the vibrations and broadening of the peaks. This
modification is most pronounced for the band at $\approx$780~cm$ ^{-1}
$, which corresponds to the symmetric stretching of the SiOSi bridges.
This band not only becomes broader but also asymmetric, with a new
peak located close to 850~cm$ ^{-1} $, a frequency which coincides with
the one of the characteristic peaks of esBO and esSi discussed in the
context of Fig.~\ref{fig:ab nsx-VDOS-es-cs}. Thus we can conclude that
the IR spectra can indeed reveal the presence of 2M rings in the sample. However, we also note that at \(\omega \approx 700\)~cm$ ^{-1}$, we find no marked peak in  \(\epsilon_2\), i.e. the peak we find at this frequency in the VDOS (see Fig.~\ref{fig:ab nsx-VDOS-es-cs}b) seems not to be IR active.

In order to understand the dependence of the spectrum on the composition
we present in Fig.~\ref{fig:ir-spectra}(b) the calculated imaginary part
of the dielectric function for the three sandwich samples. Firstly we
notice for the NS5 and NS3 glasses the presence of a broad band below
300~cm$^{-1}$, with an intensity that growth with the concentration
of Na. This trend is in agreement with experimental IR studies for bulk
glasses~\cite{merzbacher1988structure,kapoutsis1994alkali,ingram2000origins}
and a comparison with the VDOS from Fig.~\ref{fig:ab nsx-VDOS-tot-partial}
shows that this band is indeed directly related to the vibrational
motion of the sodium atoms. In contrast to this the pronounced peak at
around 400~cm$^{-1}$ depends only weakly on the concentration of sodium,
a result due to the fact that  that rocking motions of SiOSi bridges, IR active modes, are not much affected by the Na presence~\cite{kilymis_vibrational_2019}.
A stronger dependence on the Na concentration is observed for
the band from 700 to 900~cm$ ^{-1} $ in that it shifts significantly
to lower frequencies, becomes more intense, and slightly broader. The
softening of this spectral region with the addition of Na has also been
seen in experimental IR spectra for bulk glasses and attributed to
the increasing depolymerization of the network, in agreement with our
observations for our sandwich samples (see Sec.~\ref{sec:structure}).
Regarding the 2M rings we recall that their concentration decreases with
increasing Na content, accompanied by a decreasing signal in the VDOS at
$\approx$850~cm$ ^{-1} $, see Fig.~\ref{fig:ab nsx-VDOS-es-cs}. Panel~(b)
shows that at this frequency the systems with sodium do not show any sign
of a peak, i.e.~for such glasses IR spectroscopy experiments cannot be
expected to detect the presence of 2M rings in this frequency range.
Finally we mention that in the high-frequency region the addition of
Na leads to an broadening of the band and a shift of the peak to lower
frequencies.  These modifications are the signature of the increasing
number of NBOs, and they are consistent with the changes reported in
experimental
 works~\cite{merzbacher1988structure,kapoutsis1994alkali,ingram2000origins}

\section{Electronic properties}\label{sec:electronic}

In this section we present the electronic properties of our samples,
i.e.~the electronic density of states (eDOS), Bader charges, and the
electron localization function (ELF). The presence of a surface in
combination with the Na addition makes that these properties change
significantly with respect the ones for bulk silica and we will discuss
these modification in connection with the defective structures such as
2M rings or NBO.

\subsection{Electronic density of states}

\begin{figure}[t]
\centering
\includegraphics[width=0.47\textwidth]{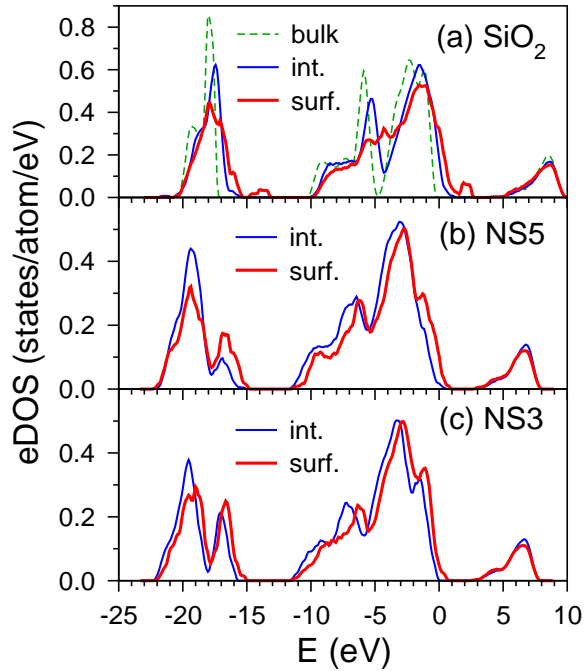}
\caption{Electronic density of states of the sandwich glasses at
0~K. Panels (a), (b) and (c) are for silica, NS5, and NS3, respectively. The
eDOS of the sandwich glasses are decomposed with respect to the surface
(surf.) and interior (int.) layers. Panel (a) shows also the eDOS for 
bulk silica. All distributions are normalized with respect to
the number of atoms. The Fermi level energy $E_f$ is at 0~eV. \\ 
}
\label{fig:ab nsx-eDOS-layers}
\end{figure}

The eDOS, $D(E)$, can be obtained directly from the Kohn-Sham
energies calculated for the structure relaxed at $T=0$~K, see
Ref.~\cite{pedesseau_first-principles_2015-1} for details.
Figure~\ref{fig:ab nsx-eDOS-layers} shows the eDOS for the interior
and surface domains of the studied compositions. For the sake of
comparison, we include in panel (a) also the data for a bulk silica
glass (dashed line), computed using the same structural model as the VDOS
discussed in the previous section~\cite{SI-commpriv}. For this bulk
system we recognize features that have been document in the literature
before~\cite{sarnthein_model_1995,benoit_model_2000}: (i) The states
at around $-20$~eV are O~$2s$ states; (ii) The states from $-10$ to $-4$~eV 
are bonding states between Si~$sp^3$ hybrids and (mainly) O~$2p$
orbitals; (iii) The states above $-4$~eV up to the Fermi level ($E=0$~eV)
are O~$2p$ nonbonding orbitals. The estimated band gap is found to
be around 5~eV, in good agreement with previous \textit{ab initio}
calculations~\cite{sarnthein_model_1995,benoit_model_2000,du_structure_2006}.
Here we mention that in general DFT calculations underestimate
the experimental band gap of materials and our result confirms
this flaw since the experimental value of the gap for silica is
9~eV~\cite{himpsel_inverse_1986,grunthaner_chemical_1986}.

For the interior layer of silica sandwich, one recognizes from
Fig.~\ref{fig:ab nsx-eDOS-layers}(a) that its eDOS is very similar to
the one of the bulk model. The main difference is that some of the peaks
are less sharp and that the main bands are shifted by around 1~eV to
higher energies. These results might be attributed to the protocol used
to prepare the samples (sandwich geometry, quench rate).  (Glass produced
with a lower cooling rate are likely to be at a lower energy state.) No
difference is found in the high energy band which makes that the band
gap for the sandwich geometry is reduced to 4.1~eV.

The eDOS' for the interior layer of the NS5 and NS3 sandwich samples,
presented in Fig.~\ref{fig:ab nsx-eDOS-layers}(b) and (c), are quite
similar to the one for silica. Certain features do, however, depend on the
composition: 1) The eDOS shifts to lower frequencies when Na is added.  2)
The splitting between O~$sp-$Si~$sp^3$ bonding and anti-bonding states is
washed out.  3) The lowest energy band has a new peak at around $-17$~eV,
and its intensity grows if the Na content increases. Below we will see
that this peak is related to the electronic states of NBO and we will also
discuss the connection of other features with structural properties.
Note that the shift in the energy scales makes that the band gaps shrink
with respect to the values of silica: We find $ 2.9 $~eV for NS5, and $
2.7 $~eV for NS3. The latter two values are also compatible with the
calculated band gap ($ 2.8 $~eV) for sodium tetrasilicate glass (i.e.,
20 mol-\% of Na$_2$O)~\cite{ispas_structural_2001}.

The eDOS for the surface layers do not differ strongly from their
counterpart for the interior layer. The distributions at negative energies
are shifted to slightly higher energies, by about 1~eV, an effect that is
likely related to the defective structures on the surface. In addition
we find that the height of the peaks is modified, notably the ones at
the lowest energies, i.e.~the O 2$s$ states, a result that is reasonable
since in the outermost layer the structure of oxygen is quite different
from the ones inside the bulk (see Fig.~\ref{fig:ab nsx-density-numfrac}.)
Finally we mention that for the case of silica the splitting between
O~$sp-$Si~$sp^3$ bonding and anti-bonding states has vanished for the
surface layer, i.e.~for these energies the eDOS is now very similar to the
one of the systems with sodium, an effect that is likely related to the increased 
structural disorder.

\begin{figure*}[t]
\centering
\includegraphics[width=0.95\textwidth]{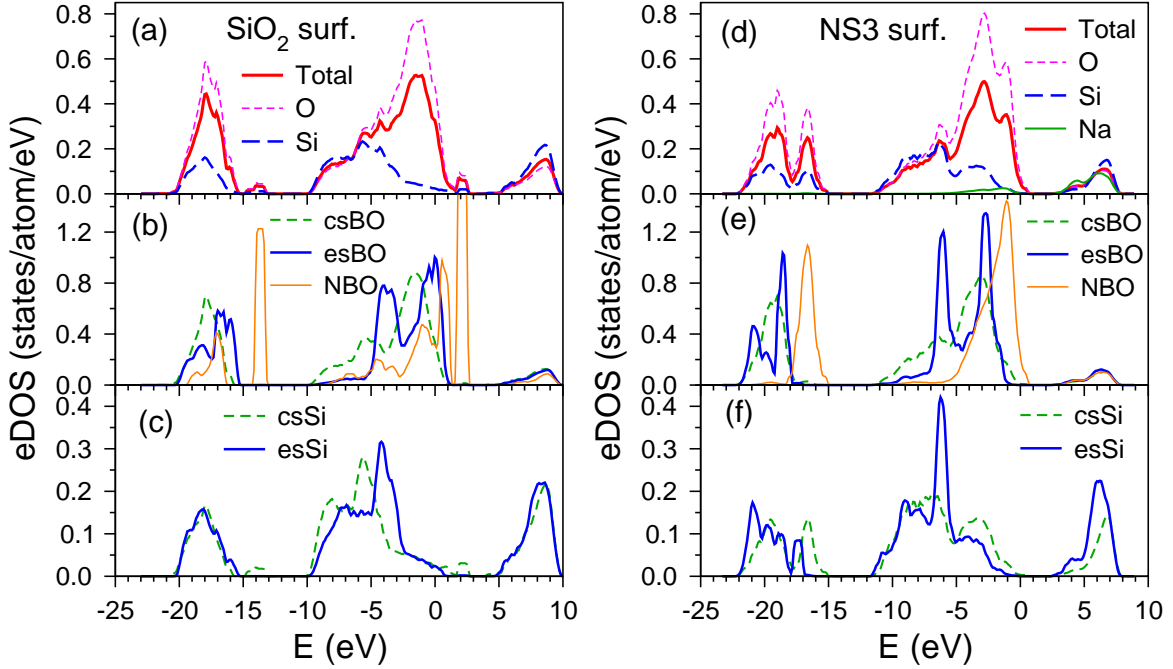}
\caption{Decomposition of the surface eDOS of the silica and NS3
glasses. Left panels: Silica. Right panels: NS3. The Fermi level energy
$E_f$ is at 0~eV. (a) and (d): Decomposition with respect to atomic
species, i.e., Si, O and Na. (b) and (e): Decomposition of O into NBO,
csO and esO. (c) and (f): Decomposition of Si into csSi and esSi.
The eDOS are normalized with respect to the number of atoms. 
}
\label{fig:ab nsx-eDOS-partial-surf}
\end{figure*}

To get insight into the relationship between the atomic structure and the
electronic properties of the glasses, we have  decomposed the eDOS of the
surface layers into partial contributions from the constituent atoms,
i.e.~Si, O, and Na, see Fig.~\ref{fig:ab nsx-eDOS-partial-surf}(a)
and (d). Subsequently we have decomposed the eDOS of Si atoms into
contributions from csSi and esSi atoms and the one of the O~atoms into
csBO, esBO and NBO atoms, i.e. the species we have found to be relevant
to characterize the structural properties of the samples, panels (b),
(c), (e), and (f).  Figure~\ref{fig:ab nsx-eDOS-partial-surf}(a) and (b)
shows that the Si and O contribute both to the band at lowest energy,
but that the distribution per atom is about 3 times larger for O than for
Si. For the energies between $-10$~eV and $-5$eV both species have a very
similar density, as it is the case for the states with positive energy,
but that the band between $-5$~eV and the Fermi energy the signal is
strongly dominated by oxygen. These results hold also for the case of
NS3, panel (d), since we see that Na contributes basically only to the
band at positive energies, i.e.~the conduction band.

The further decomposition of the eDOS for the silica surface shows that
the two small peaks at around $-14$ and 2~eV in the total eDOS are mainly
due to states of NBO atoms, see Fig.~\ref{fig:ab nsx-eDOS-partial-surf}(b)
and (e), and with a weak contribution from csSi atoms, panel (c).  Therefore, these two peaks can be assigned to Si-O dangling bonds,
in agreement with the findings of previous first principles simulations
for hydrated silica~\cite{benoit_nature_2008}.  These NBO give also rise
to a signal at around $-17$~eV which makes that the total eDOS for silica
has a shoulder at around this energy, panel (a), and the one for NS3 a
pronounced peak, panel (d).

For the silica surface, we note that the main valence band for the
edge-sharing atoms is shifted by about 2~eV to higher energies, panel
(b). This shift makes that the peaks and valleys in the distributions for
the csBO and esBO cancels each other, resulting in a total distribution
that is rather featureless, i.e.~the splitting between O~$2p-$Si~$sp^3$
bonding and O~$2p$ nonbonding states in the total eDOS of silica surface
has disappeared, panel (a). The atoms of the 2M rings, i.e.~esSi and esBO, give rise
to peaks  between $-20$ and $-15 $~eV  and  $-10$ and $ 0 $~eV,
features that are consistent with DFT calculation for cristalline fibrous silica
containing these particular defective structure~\cite{hamann1997energies}. 

Comparing panels (b) and (c) for silica with the corresponding ones
for NS3, panels (e) and (f), one sees that the various distributions
are quite similar. The main difference is that the ones for NS3 are
slightly shifted to lower energies. Hence we can conclude that the
corresponding shift with sodium concentration, made already in the context
of Fig.~\ref{fig:ab nsx-eDOS-layers}, is due to the shift of the energies
of the individual species.

\subsection{Bader charges}
Further insight into the electronic properties of the glasses can be
obtained by analyzing how  the charge density can be assigned to the various
type of atoms. 
To this aim we have employed the ``atom in molecule''
(AIM) approach proposed by Bader~\cite{bader_atoms_1985}, which allows to
partition the electron density $\rho(\*r)$ among the constituent atoms
and thus to define the atomic charges. The Bader
charge is given by

\begin{equation}
Q_\alpha^{\rm Bader} = Z_\alpha -\int_{V_{\rm Bader} } \rho({\bf r})dV,
\end{equation}

\noindent
where $Z_\alpha$ is the number of valence electrons of an atom $\alpha$
and $V_{\rm Bader}$ is the so-called Bader volume around the atom.
By definition, the Bader volume is limited by a
surface $S({\bf r})$ which exhibits a zero flux property, i.e., the inner
product $\nabla\rho({\bf r})\cdot{\bf n}=0$, where ${\bf n}$ is the unit
vector oriented perpendicular to $S({\bf r})$~\cite{bader_atoms_1985}.

In Tab.~\ref{tab: ab bader-charge} we list the average Bader charges of
various atomic species in the three glasses. Note that, in contrast to
the structural analysis, for the charge analysis we did not distinguish
between the surface and interior layers since we found no significant
difference between the two. This suggests that the Bader partition scheme does not allow to establish direct relationships with the different structural and vibrational properties of the surface and interior domains (discussed in the previous sections).

\begin{table*}[ht]
\small
\center
\begin{tabularx}{10cm}{lYYY}
\hline
		Charge ($e$) & Silica        & NS5           & NS3           \\ \hline
		Si         & 3.154(0.151)  & 3.150(0.106)  & 3.146(0.025)  \\
		Si$^{3}$        & 2.458(0.469)  & -          & -             \\
		Si$^4$        & 3.176(0.018)  & 3.156(0.024)  & 3.146(0.025)  \\
		Si$^5$        & 3.201     & 3.178(0.009)  & 3.158(0.027)  \\  \hline
		$Q_2$         & 3.136         & 3.113(0.007)  & 3.105(0.017)  \\
		$Q_3$         & 3.142(0.009)  & 3.133(0.018)  & 3.134(0.015)  \\
		$Q_4$         & 3.177(0.018)  & 3.169(0.016)  & 3.168(0.015)  \\  \hline
		O          & -1.577(0.08)  & -1.586(0.055) & -1.588(0.03)  \\
		NBO        & -1.106(0.243) & -1.529(0.071) & -1.543(0.01)  \\
		BO         & -1.587(0.014) & -1.599(0.012) & -1.606(0.011) \\  \hline
		esBO       & -1.563(0.009) & -1.584(0.012) & -1.586(0.011) \\
		esSi       & 3.144(0.013)  & 3.13(0.022)   & 3.119(0.025)  \\  \hline
		Na         & -             & 0.847(0.015)  & 0.84(0.016)   \\ \hline
\end{tabularx}
\caption{
\label{tab: ab bader-charge} 
Average Bader charge of atoms and various species found in the three
glasses at 0~K. The values given in parentheses are the standard deviation of
their distributions.  No values in parenthesis means that only one such
specie has been found. 
}
\end{table*}

For Si, the average charge of Si$^4$ (i.e.,~an Si bonded to four O) in
the silica glass is about +3.18~$e$, in good quantitative agreement with
the result found in quartz (+3.20~$e$)~\cite{gibbs_model_1999}, in bulk
amorphous silica~\cite{pasquarello1997dynamical}, for a silver/silica
interface~\cite{balout2019density} or for $\beta$-cristobalite
surfaces~\cite{le2018structural}.  In addition, we find that
$q_{\rm Si}$ increases with increasing coordination number $n$,
see rows Si$^n$, in qualitative agreement with observations from a high-energy
synchrotron-radiation study of stishovite (the high-pressure polymorph
of silica)~\cite{kirfel_electron-density_2001}. Furthermore we note
that $q_{\rm Si}$ depends also on the character of the tetrahedron,
$Q_m$, where $m$ denotes the number of BO connected to the Si atom,
in that $q_{\rm Si}$ increases with $m$. By comparing the Si charge
of the three glasses, one notices that $q_{\rm Si}$ decreases with
increasing Na concentration. An inspection of the Na-dependence of the
$Q_m$ species shows that this decrease is likely due to the change in
the concentration of the $Q_n$ species and not to the Na-dependence of
their charge, since the latter is rather weak.

For oxygen we find that the average charge of BO is close to $-$1.59~$e$,
a value which is in agreement with the one obtained for $\alpha-$quartz,
$-1.60~e$~\cite{gibbs_model_1999}, and other systems containing silicon and
oxygen~\cite{pasquarello1997dynamical,balout2019density,le2018structural}.
The table also shows that $q_{\rm BO}$ is more negative than $q_{\rm
NBO}$, a deficiency of the Bader charge analysis which has already
been found in previous \textit{ab initio} simulations, see for example
Refs.~\cite{du_structure_2006,pedesseau_first-principles_2015-1}. Despite
this flaw, it is still instructive to discuss the atomic charges
in different systems using the same description. Table~\ref{tab:
ab bader-charge} shows, e.g., that the $q_{\rm O}$ becomes slightly
more negative with the addition of Na. This trend is mainly due to the
pronounced Na-dependence of the charge of the NBO.

Regarding the 2M~rings, we find that the esSi atoms are slightly less
charged than the average Si atoms. This can be rationalized by the fact
that in 2M~rings the two oxygen atoms are quite close to each other
which makes that their electron clouds are pushed in the direction
of the Si atoms, making that the charge of the latter decreases. This
interpretation is coherent with the observation that the esBO have a
charge that is less negative than the one of the ordinary BO.

Finally, we note that the Na charge has a value of $\approx+$0.84~$e$
and is basically independent of the Na concentration. This result is in
good quantitative agreement with a previous \textit{ab initio} simulation
of a sodium borosilicate glass, where a Bader charge of +0.83~$e$ was
found for Na ions~\cite{pedesseau_first-principles_2015}.

\subsection{Electron localization function}

\begin{figure*}[htp]
\center
\includegraphics[width=0.91\textwidth]{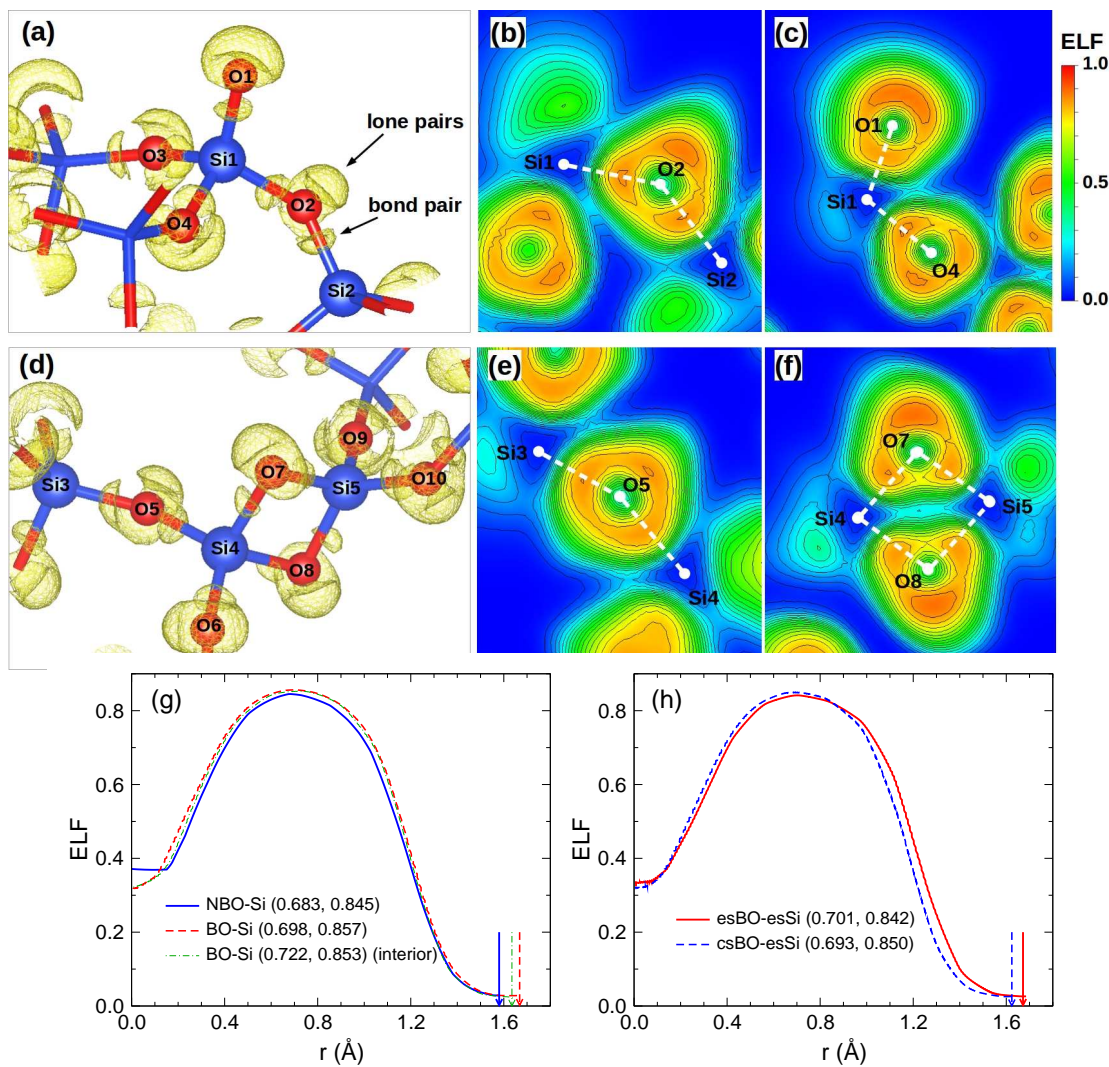}
\includegraphics[width=0.81\textwidth]{fig12gh-ELF-ns0-surface-2d-contour-plot_new.eps}
\caption{Analysis of chemical bonding on a SiO$_2$ surface by means of the
electron localization function (ELF). (a) The representation of the ELF for a small region
on the surface highlighting a SiO$_4$ tetrahedron, centered on an
Si atom labeled Si1, bonded to one NBO, O1, and three BO atoms (O2-O4).
The iso-surface (in yellow) corresponds to the ELF at a value of 0.83.
(b) and (c): 2D contour plots of the ELF in the planes defined by three
atoms: Si1-O2-Si2, panel (b), and O1-Si1-O4, panel (c), where the atoms are identified in panel (a).
The increment of iso-lines is 0.05. (d)-(f): The same
representation as in (a)-(c) but for a two-membered ring structure. (g) and
(h): Line profiles of the ELF along the bond paths as shown in (a)
and (d), respectively.  Also included in (g) is the average ELF profile
of the BO-Si bonds that belongs to a Si-BO-Si connection in the interior
domain (green dash-dotted line). The O atom is at $r=0$. For each
bond path the point corresponding to the maximum ELF is indicated in the
parenthesis ($r$, ELF$(r)$). The arrows show the location of the average Si-O bond length.  The
visualization of the ELF was realized by VESTA~\cite{momma_vesta_2011}. 
}
\label{fig:ab elf-ns0-surf}
\end{figure*}

In this subsection we discuss the nature of the chemical
bonding in the glasses using the electron localization function
(ELF)~\cite{becke_simple_1990}. The ELF is related to the probability
distribution $\eta({\bf r})$ of electron pairs, divided by the
corresponding distribution for a uniform electron gas. By definition,
$\eta$ takes at any point of space a value that lies between 0 and
1. A value of 1 corresponds to a perfect localization of the electron
pairs, while a value of 0.5 corresponds to that of a uniform electron
gas. Details of the calculation can be found in Ref.~~\cite{savin_elf:_1997}.

In Fig.~\ref{fig:ab elf-ns0-surf} we illustrate some of the properties
of the ELF for the case of the silica glass surface. Panel (a) shows the
iso-surface of the distribution evaluated at the value $\eta=0.83$. The
region we consider includes a SiO$_4$ tetrahedron with one NBO (marked
as O1) and three BO (O2-O4). For each BO  we observe a  hemispherical
domain along each Si-O bond, see for example the bridge Si1-O2-Si2 in
panel (a), and this domain can be assigned to a
pair of bonding electrons. One also finds a banana-shaped domain
at the reflex side of the Si-BO-Si bridge, which is orthogonal  to
the Si-BO-Si plane. This domain is assigned to two lone pairs of
electrons, i.e., the four valence electrons that are not involved in
bonding. These non-bonding domains  are substantially
larger than the bonded hemispherical domains along the Si-O bonds,
in agreement with the ELF mapping of the SiOSi linkage in silicate
minerals~\cite{gibbs_mapping_2005}. For the NBO atoms, as for example
the atom labelled O1, we observe that, besides the bond pair domain,
a concave hemispherical-shaped domain can be found and it seems to
have a rotational symmetry along the Si-NBO direction. This domain
can be ascribed to the non-bonding electrons and it appears to have
a larger volume than the nonbonding domain electron domain for BO.
This observation is reasonable since presumably there are five nonbonding
electrons for the NBO while only four for the BO.

Figure~\ref{fig:ab elf-ns0-surf}(b) shows the two-dimensional contour
plot of the ELF in a plane spanned by Si1, O2, and Si2, i.e.~for a BO,
and, panel (c), for the plane given by O2, Si1, and O1, i.e.~for a NBO.
The aforementioned bonding and nonbonding domains are clearly visible
from the contour plots. In addition, one recognizes from panel (c) that
the probability distribution of electron pairs around the NBO is more
spread out than that of the BO. This observation can be rationalized by
the fact that the NBO has more free volume on the side opposite to the
Si-O bond than the BO atoms.

A further important structural unit, namely a 2M ring, is depicted in
Fig.~\ref{fig:ab elf-ns0-surf}(d). One notices that the O atoms in the
2M ring, O7 and O8, have electron pair domains that are similar to the
ones of ordinary BO atoms, e.g., O2 in panel (a). Figure~\ref{fig:ab
elf-ns0-surf} (e) and (f) show the ELF contour plots corresponding to
two Si-O-Si linkages associated with the 2M ring. (Note that the Si-O-Si
linkage in panel (e) involves an edge-sharing Si, Si4.) One sees that the
angle Si3-O5-Si4 is much larger than the one in panel (b), demonstrating
that the strong angular constraint in the 2M ring also affects the
linkages of its neighbors. Consequently, the bond and lone pair domains
around the BO in panel (e) are not as well structured as the ones in
panel (b).  Panel (f) shows the ELF contour plots of the 2M ring. One
observes that the bond and lone pair domains are well structured and
can be clearly distinguished.  Another noticeable feature is that the
bond paths, i.e. the lines connecting neighboring atoms, are no longer
axes of symmetry for the bond pair domains. This is likely due to the
strong repulsion of the electrons from the two opposing esBO atoms.

To describe the ELF in a more quantitative manner we show in
Fig.~\ref{fig:ab elf-ns0-surf}(g) and (h) the line profile of the
ELF along the bond paths starting from the oxygen atom ($r=0$). Note
that all BO in panel (a) and (g) are ordinary corner-sharing BO,
i.e.~csBO. Figure~\ref{fig:ab elf-ns0-surf}(g) shows that the ELF of the
NBO-Si bond is smaller than the one of the BO-Si bond, implying that the
ELF around the NBO is more spread out, in agreement with the contour
plot in panel (c). In addition we note that the BO-Si bond peaks at a
larger $r$ that the Si-NBO bond (see the values in the parentheses of the
legend), in agreement with the observation that for the NBO the ELF
is extended in the direction opposite to the Si-O bond. Also included in
panel (g) is the ELF profile corresponding to a Si-BO-Si linkage in the
interior of the sample and which has an angle close to the Si1-O2-Si2
linkage shown in panel (a). 
The presence of the surface does not seem to affect in a significant manner the ELF profile of the Si-BO bonds, although the BO-Si bond length (indicated by the vertical arrows) in the
interior is slightly smaller than the surface BO-Si (see also Tab. 
~\ref{tab:ab-bond-liquid-glass}). 
Figure~\ref{fig:ab elf-ns0-surf}(h) compares 
the ELF line profiles of the esBO-esSi and csBO-esSi bonds and one notices
that the ELF of the esBO-esSi bonds shifts to a larger $r$ relative to
the csBO-esSi bonds but seems to have the same maximum height.  However,
since for the esBO-esSi bond the bond path does not pass through the
maximum of the ELF, see panel (f), the real maximum value of the ELF
for this bond is in fact higher than the one for the csBO-esSi bond, i.e. the electrons are
more localized.

\begin{figure*}[htp]
\includegraphics[width=0.95\textwidth]{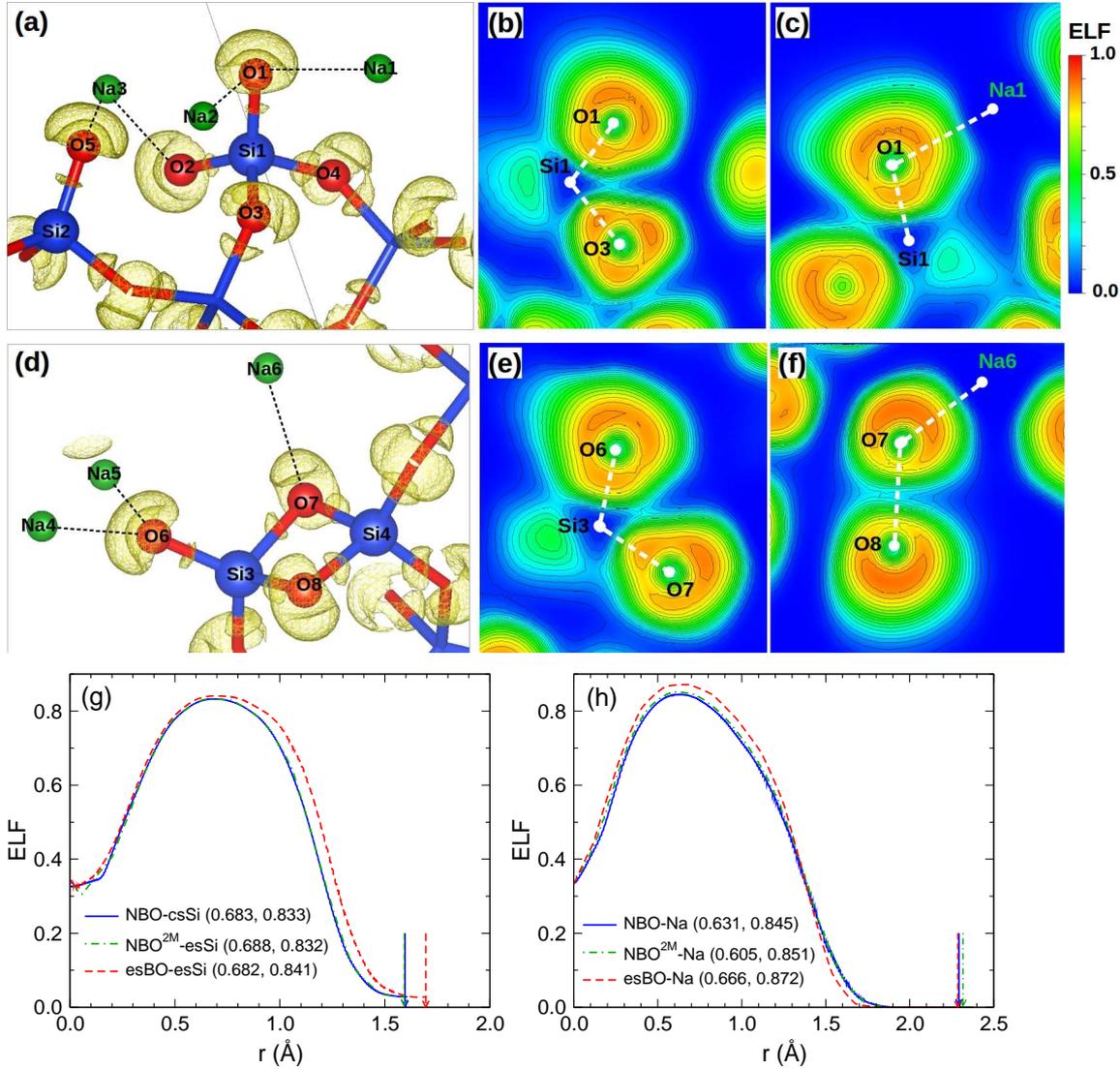}
\includegraphics[width=0.85\textwidth]{fig13gh-ELF-ns3-surface-2d-contour-plot_new.eps}
\caption{Analysis of chemical bonding on the surface of NS3 by the electron localization
function (ELF). (a) A map of the ELF for the structures
on the surface, highlighting a SiO$_4$ tetrahedron which has a Na in its neighborhood. The dashed lines are the O-Na bonds with $r_{\rm
O-Na}<2.5$~\AA. The iso-surface represents the ELF surface at a value of 0.83 and the
assignment of different domains are the same as in Fig.~\ref{fig:ab elf-ns0-surf}. (b)
and (c): 2D contour plots of the ELF in the planes defined by three
atoms. The increment of iso-lines is 0.05. (d-f): The same
representation as in (a)-(c) but for a two-membered ring structure. (g) and
(h): Line profiles of the ELF along the bond paths as shown in
(a) and (d), respectively. The oxygen atom is at $r=0$. For each bond path the
point corresponding to the maximum ELF is indicated in the parenthesis. The
arrows show the location of the average Si-O or Na-O bond length. NBO$^{\rm 2M}$
denotes the NBO bonded to an esSi.  The
visualization of the ELF was realized by VESTA~\cite{momma_vesta_2011}. 
}
\label{fig:ab elf-ns3-surf}
\end{figure*}

Figure~\ref{fig:ab elf-ns3-surf} shows the ELF results for the NS3
glass surface. We note that, in addition to the structural modification
discussed in the previous sections, the presence of Na induces also changes
in the bonding. For example, panel (a), the bond pair domain for the
NBO-Si bond O1-Si1 is much smaller that the corresponding domain in silica,
Fig.~\ref{fig:ab elf-ns0-surf}(a). Figure~\ref{fig:ab elf-ns3-surf}(b)
shows that the presence of Na also leads to an asymmetry of the lone
pair domain of the NBO (i.e.~O1). This effect is also seen from the
two dimensional contour plot in the plane defined by Na1-O1-Si1, panel (c).
For the NBO, O1, we note
that the domains in the directions of the Na atoms can be ascribed to
the Na-O bond pair interaction superimposed on the lone pair domains,
panel (c). Similar results were found for earth materials containing alkali
	metals~\cite{gibbs_mapping_2005}.

Figure~\ref{fig:ab elf-ns3-surf}(d) shows a 2M ring  with one of the Si atoms connected to a
NBO and its nearby Na atoms.
Panel (e) shows that, for the 2M rings,
the distribution is no longer symmetric around the O7(esBO)-Si3 connection, an
observation that is coherent with the finding for the 2M rings in silica,
see Fig.~\ref{fig:ab elf-ns0-surf}. For the  NBO, O6, we find that the ELF contour plot  is quite similar to the one for O1 shown in panel (b), in spite of the presence of the  neighboring 2M ring.
 Figure~\ref{fig:ab elf-ns3-surf}(f)
clearly shows that the ELF for the esBO (O7) bonded to the Na is less spread out
than the distribution for the other esBO (O8) in the 2M ring, demonstrating that O7 is indeed
bonded to the Na atom.

Figure~\ref{fig:ab elf-ns3-surf}(g) shows the average ELF line profiles of
various types of O-Si bonds. (Note that the NBO atom connected to an esSi
atom is denoted as NBO$^{\rm 2M}$.) One observes that the ELF profile of
the NBO-csSi bond is very similar to the one of the NBO$^{\rm 2M}$-esSi
bond, indicating that the NBO-Si bond character is basically independent
of the Si type. Furthermore we find that the ELF values of the NBO-Si
bonds are smaller than that of the esBO-esSi bond, in accordance with
the fact that the distribution of the electron pairs around the NBO
is more spread out than the one for the esBO-esSi bond. (Also here we
recall that the ELF for the esBO-esSi is not symmetric with respect to
the connecting axis, see panel (c), and hence the maximum value is even
higher.) For all the three NBO-Si bonds, the maximum of the ELF is located
at $r\approx0.68$~\AA, independent of the bond type. Figure~\ref{fig:ab
elf-ns3-surf}(h) shows the profiles for the O-Na pairs and one sees 
that the maxima of the curves are located
at $r\approx0.61$, 0.63 and 0.67~\AA\ for the NBO$^{\rm 2M}$-Na, NBO-Na,
and esBO-Na bonds, respectively. This results indicate that the character
of the O-Na bond is more sensitive to the changes in local environment
than the NBO-Si bond. We also note that the maxima of the ELF for the O-Na
bonds are closer to the oxygen atoms (at $r=0$) than the ones of the O-Si
bonds. This result demonstrates that the O-Na is less covalent (i.e. more
ionic) than the O-Si bonds. In addition, based on the locations of the
ELF maxima, it can be deduced that the esBO-Na bond is more covalent
than the NBO-Na bonds.

Finally, we note that the locations of the maxima of the ELF profiles
for the NBO-Si and esBO-esSi bonds are very close to the corresponding
values found for the silica glass. This similarity indicates that the
presence of Na affects the position of the bond pair domains of the O-Si
bonds only weakly.

\section{Summary and Conclusions} \label{sec:conclusions}

Using {\it ab initio} calculations, we have studied the structural,
vibrational, and electronic properties of the surface of amorphous silica
and two binary sodo-silicate glasses. Previous studies have shown that,
for the case of silica, two-membered rings are an important structural
motif at the surface~\cite{rimola_silica_2013,tielens2019Characterization}. 
The present analysis of the compositional
dependence of the surface and interior domains of our sandwich samples
shows that the concentration of defect sites is considerable reduced
with increasing Na content since sodium migrates from the interior to
the surface and transforms energetically unfavorable local structures, such as 2M rings,
into more relaxed ones. As a consequence the frequency of two-membered
rings decreases rapidly with the addition of sodium.

From the dynamical matrix of the samples we have calculated the total
vibrational density of states as well as the contributions of the various atomic
species and structural elements to this distribution. This has allowed
us to identify the spectroscopic signatures of the 2M rings and see how
these change as a function of the sodium content. In addition we have
computed the IR spectra and have also determined also for this observable
the signature of the 2M rings. 
These calculations show that not all vibrational modes of the 2M rings are IR active, thus pointing out the need to use additional experimental techniques to study these rings.
In
addition the present study can serve as a benchmark for simulations of
glass surfaces using effective potentials since our results will allow to
compare the results of the classical MD simulations with highly accurate
microscopic structural and vibrational data.

Taking advantage of the {\it ab initio} approach, we have probed the
electronic properties of the glass samples with a particular focus on
the surfaces. The analysis of the electron localization function shows
that 2M rings and NBO do have a distinct electronic distribution and we
have investigated how it is affected by the presence of sodium.
To the best of our knowledge, the current simulations and analysis
represent the first study that investigates simultaneously the structural,
spectroscopic, and electronic properties of silica glass surface and how
they evolve with Na addition.  Hence this approach allows to circumvent the
frequently encountered problem that the samples probed with different
techniques usually have different production histories (cooling
rates, composition, atmospheres, etc.) which  makes the unambiguous
identification of the various structural features difficult. As a
consequence the present work should be a relevant step forward in our
understanding of the properties of oxide glasses on a quantitative level.

\section*{Acknowledgements}
Z.Z. acknowledges financial support by China Scholarship Council (NO. 201606050112).
W.K. is member of the Institut Universitaire de France.
This work was granted access to the HPC resources
of CINES under the allocation  A0030907572, A0050907572 and  A0070907572
attributed by GENCI (Grand Equipement National de Calcul
Intensif).


\end{document}